
%
%

\overfullrule=0pt
\parskip=3pt
 at 14truept
\overfullrule 0pt


\catcode`\@=11
\font\tenmsa=msam10
\font\sevenmsa=msam7
\font\fivemsa=msam5
\font\tenmsb=msbm10
\font\sevenmsb=msbm7
\font\fivemsb=msbm5
\newfam\msafam
\newfam\msbfam
\textfont\msafam=\tenmsa  \scriptfont\msafam=\sevenmsa
  \scriptscriptfont\msafam=\fivemsa
\textfont\msbfam=\tenmsb  \scriptfont\msbfam=\sevenmsb
  \scriptscriptfont\msbfam=\fivemsb

\def\hexnumber@#1{\ifcase#1 0\or1\or2\or3\or4\or5\or6\or7\or8\or9\or
        A\or B\or C\or D\or E\or F\fi }

\def\Bbb{\ifmmode\let\next\Bbb@\else
 \def\next{\errmessage{Use \string\Bbb\space only in math mode}}\fi\next}
\def\Bbb@#1{{\Bbb@@{#1}}}
\def\Bbb@@#1{\fam\msbfam#1}

\def\\{\hfil\break}
\def\la{\lambda}
\def\half{{1 \over 2}}
\def\om{\omega}
\def\si{\sigma}
\def\Z{{\Bbb Z}}
\def\l{\ell}
\def\cqfd{\hskip 2truemm \vrule height2mm depth0mm width2mm}
\def\D{{\cal D}}

\def\al{\alpha}
\def\be{\beta}
\def\ga{\gamma}
\def\Q{{\cal Q}}
\def\b#1{\big(#1\big)}
\def\ka{\kappa}
\def\+{\oplus}
\def\x{\otimes}

\font\huge=cmr10 scaled \magstep2
\font\small=cmr8


{\nopagenumbers
\rightline{LETH--PHY--3/95}
\rightline{US--FT/22--94}
\rightline{UCL--IPT--95/4}
\rightline{March 12, 1995}
\vskip3cm
\centerline{{\huge\bf Automorphism Modular Invariants of
Current Algebras}}
\bigskip\bigskip\centerline{T. Gannon$^\bullet$\footnote{$^*$}{\small
E-mail: gannon@abacus.concordia.ca},
P. Ruelle$^\circ$\footnote{$^\star$}{\small Chercheur
Qualifi\'e FNRS. On leave from: Institut de Physique Th\'eorique, Universit\'e}
\footnote{}{\small Catholique de Louvain, Louvain--la--Neuve, Belgium.
E-mail: ruelle@fyma.ucl.ac.be} and M.A.
Walton$^\odot$ {\footnote{$^\dagger$}{\small Supported in part by NSERC.
E-mail:
walton@hg.uleth.ca}}}
\bigskip
\centerline{$^\bullet${\it Mathematics Department, Concordia University}}
\centerline{{\it Montr\'eal, Canada, H3G 1M8}}
\bigskip
\centerline{$^\circ${\it Departamento de F\'\i sica de Part\'\i culas}}
\centerline{\it Universidade de Santiago}
\centerline{\it 15706 Santiago de Compostela, Spain}
\bigskip
\centerline{$^\odot${\it Physics Department,
University of Lethbridge}} \centerline{{\it Lethbridge, Alberta,
Canada\ \ T1K 3M4}}

\vskip3cm \leftskip=2cm \rightskip=2cm
\noindent{{\bf Abstract:}}
We consider those two-dimensional rational conformal field theories (RCFTs)
whose chiral algebras, when maximally extended, are isomorphic to the
current algebra formed from some affine non-twisted Kac--Moody algebra at
fixed level. In this case the partition function is
specified by an automorphism of the fusion ring and corresponding symmetry of
the Kac--Peterson modular matrices. We classify all such partition functions
when the underlying finite-dimensional Lie algebra is simple. This gives
all possible spectra for this class of RCFTs. While accomplishing this,
we also find the primary fields with second
smallest quantum dimension.

\leftskip=0cm \rightskip=0cm

\vfill

\eject}

\pageno=1
\noindent{\bf 1. Introduction}

\vskip 0.7truecm \noindent

In two--dimensional conformal field theory, scale invariance means
boundary conditions have an impact on the local
physics, even far from a boundary [6]. For example, a conformal field
theory must be consistent on the interior of a parallelogram with periodic
boundary
conditions imposed, i.e. on a torus. In particular, the corresponding partition
function should not be sensitive to changes of the modular
parameter that keep a torus within the same conformal class. The partition
function must be modular invariant.

The local symmetry of the conformal field theory also constrains the partition
function. The chiral algebra of currents determines the conformal blocks [2]
of the torus partition function. That is, the partition function must be a
sesquilinear combination of characters of the chiral algebra. The two
constraints together often determine completely the field content of a given
conformal field theory. This analysis of conformal field theories is known as
the modular bootstrap.

We apply the modular bootstrap program to conformal field theories whose
(maximal) chiral algebras are isomorphic to the current algebra of
nontwisted affine Kac--Moody algebras at fixed levels. We call
such algebras {\it conformal current algebras}, and the corresponding theories
{\it unextended current models}. Their partition functions are described by a
permutation matrix  that also gives the action of a fusion rule automorphism
[27]. For this reason, candidates for such partition functions are known as
automorphism (modular) invariants. We will limit our attention here to the
case where the underlying finite-dimensional Lie algebra is simple.

We actually solve the slightly more general problem of finding, for each
simple Lie
algebra $X_\ell$ and level $k$, the set of all permutations $\si$ of the
alc\^ove $P_+(X_{\ell,k})$ of highest weights (see (3.1) below), which are
symmetries of the corresponding Kac-Peterson modular matrices, i.e. which
obey equations (3.8a),(3.8b) below. This is what we mean by {\it
automorphism invariants}. Their classification should be of mathematical
value independent of RCFT. The question of which of these are actually
realized as the partition function of a RCFT is not addressed here.

There are in the literature at least two different meanings of the phrase
{\it Wess-Zumino-Witten (WZW) models}. The more general one is any RCFT
whose maximal chiral algebras contain a conformal current algebra such that
any character of the former can be written as a finite sum of characters
of the latter. The partition function for such a RCFT will then be a
finite sesquilinear combination of affine algebra characters. We suggest
the term {\it current models} for these; when the chiral algebra equals
the current algebra, we will call them {\it unextended} current models. A
more restrictive definition are those RCFTs corresponding to a string
moving on a compact Lie group [18] --- we retain the term {\it WZW model}
for these. The WZW partition function has been computed for each simple,
compact, connected Lie group [13]. In this paper we find all automorphism
invariants; our list will include all possible partition functions for
the unextended current models.
Many of these automorphism invariants
cannot be found in [13], and some still lack such an explicit interpretation.

The list of partition functions of unextended current models is presented
below, and proved complete. This result is a major step towards the more
ambitious classification of {\it all} modular invariants of current models,
including those described by a chiral algebra that extends the conformal
current algebra. The list we find is also the useful one from the point of view
of symmetry. It is often easier to identify the symmetry of a physical theory,
before identifying the details of the dynamics. In that sense, a list of
possible partition functions with a given maximal chiral algebra is the most
relevant. Our catalogue gives the complete list for each (simple) conformal
current algebra.

This work follows [17], where the automorphism invariants for
algebra $A_{m_1} \oplus\cdots\oplus A_{m_s}$
were treated. Our restriction here to simple Lie algebras is convenient,
but as [17] shows, the generalization to semi-simple Lie algebras should
be possible. It is hoped that classification results like ours will teach us
something about more general classes of conformal field theories, perhaps all
rational ones. The greatest impetus to this program was given by the curious
A--D--E classification of $A_1$ modular invariants [5]. Extension of this
work proved difficult: the $A_2$ invariants were only recently
classified in [16] (special cases of the $A_2$ classification were also
obtained  in [29]). Because we treat all simple Lie algebras here, our
work may reveal new features of these modular invariants families  that are
universal
(previously, only the level one theories had been classified for all simple
Lie algebras [21,15]).

Our results are stated in section 2, along with a brief outline of the
classification proof. Sections 3 through 7 are devoted to the proof. A short
conclusion is given in section 8.

\vskip1truecm
\noindent{\bf 2. Statement of the Results}

\vskip 0.7truecm \noindent
The paper will be devoted to the proof of the following statement, already
proved  for the $A_\l$ series in [17].

\bigskip \noindent
{\bf Theorem.} {\sl The complete list of automorphism invariants $\si$ for
the current algebra $X_{\l,k}$, where $X_\l$ is a simple Lie algebra and
$k\in{\Bbb N}$, is given in Tables 1 and 2.}

\vskip 0.5truecm \noindent
Explicit definitions of all automorphism invariants listed in Tables 1 and 2
are given by the relevant subsections of sections 5,6 and 7. The total
number of automorphism invariants for fixed $X_{\l,k}$ is given in the
third column. These form a group under composition $\si\circ\si'$; this
group is given in the final column of the tables.

{}From Tables 1 and 2, we see that the automorphism invariants can all be
described solely in terms of the symmetries of the extended Dynkin
diagram (conjugations and simple currents --- see section 3), with the
exceptions of $B_{\l,2}$,
$D_{\l,2}$, $E_{8,4}$, $F_{4,3}$ and $G_{2,4}$, where exceptional
automorphism invariants appear. Except for the $E_{8,4}$ one, these
exceptionals stem
from Galois transformations, with a subtle touch of simple
currents --- this will be discussed in more detail in section 3.

\topinsert
\centerline{
{\vbox{\offinterlineskip
\hrule
\halign{& \vrule# & \strut \quad \hfill#\hfill \quad
& \vrule# & \strut \quad \hfill#\hfill \quad
& \vrule# & \strut \quad \hfill#\hfill \quad
& \vrule# & \strut \quad \hfill#\hfill \quad
& \vrule# & \strut \quad \hfill#\hfill \quad & \vrule# \cr
height5pt & \omit && \omit && \omit && \omit && \omit &\cr
& $X_{\l,k}$ && conditions && \# autom. && names && group & \cr
height5pt & \omit && \omit && \omit && \omit && \omit &\cr
\noalign{\hrule}
height7pt & \omit && \omit && \omit && \omit && \omit &\cr
&$A_{\ell,k}$&&&&$2^{c+p+t}$&&$\{C^a\si_m\}$&&${\Bbb D}_1^{c+p+t}$&\cr
height7pt & \omit && \omit && \omit && \omit && \omit &\cr
\noalign{\hrule}
height7pt & \omit && \omit && \omit && \omit && \omit &\cr
&$B_{\ell,1}$&&&& 1 &&$\{\si_1\}$&&&\cr
height4pt & \omit && \omit && \omit && \omit && \omit &\cr
&$B_{\ell,2}$&&&&$2^{p-1}$&&$\{\si_m\}$&&${\Bbb D}_1^{p-1}$&\cr
height4pt & \omit && \omit && \omit && \omit && \omit &\cr
&$B_{\ell,k}$&&$k\ge 3,\ k\ {\rm odd}$&&$2$&&$\{\si_1,\si_J\}$&&${\Bbb D}_1$&
\cr
height4pt & \omit && \omit && \omit && \omit && \omit &\cr
&$B_{\ell,k}$&&$k\ge 4,\ k\ {\rm even}$&&$1$&&$\{\si_1\}$&&&\cr
height7pt & \omit && \omit && \omit && \omit && \omit &\cr
\noalign{\hrule}
height7pt & \omit && \omit && \omit && \omit && \omit &\cr
&$C_{2,1}$&&&& 1 &&$\{\si_1\}$&&&\cr
height4pt & \omit && \omit && \omit && \omit && \omit &\cr
&$C_{\ell,k}$&&$k\ell \equiv_4 2,\ (\ell,k)\ne (2,1)$
&&2&&$\{\si_1,\si_J\}$&&${\Bbb D}_1$&\cr
height4pt & \omit && \omit && \omit && \omit && \omit &\cr
&$C_{\ell,k}$&&$k\ell \not\equiv_4 2$&&$1$&&$\{\si_1\}$&&&\cr
height7pt & \omit && \omit && \omit && \omit && \omit &\cr
\noalign{\hrule}
height7pt & \omit && \omit && \omit && \omit && \omit &\cr
&$D_{\ell,1}$&&$\ell\equiv_8 4$&&$6$&&$\langle\si_s,\si_c\rangle$&&${\Bbb
D}_3$&\cr
height4pt & \omit && \omit && \omit && \omit && \omit &\cr
&$D_{\ell,1}$&&$\ell\not\equiv_8 4$&&$2$&&$\{\si_1,C_1\}$&&${\Bbb D}_1$&\cr
height4pt & \omit && \omit && \omit && \omit && \omit &\cr
&$D_{4,2}$&&&&$6$&&$\{C_j\}$&&${\Bbb D}_3$&\cr
height4pt & \omit && \omit && \omit && \omit && \omit &\cr
&$D_{4,k}$&&$k>2,\ {\rm even}$&&$12$&&$\{C_j\si_{vsc}^a\}$&&${\Bbb D}_6$&\cr
height4pt & \omit && \omit && \omit && \omit && \omit &\cr
&$D_{4,k}$&&$k>1,\ {\rm
odd}$&&$36$&&$\{C_j\langle\si_s,\si_c\rangle\}$&&${\Bbb D}_3^2$&\cr
height4pt & \omit && \omit && \omit && \omit && \omit &\cr
&$D_{\ell,2}$&&$\ell>4$&&$2^p$&&$\{C^a_1\si_m\}$&&${\Bbb D}_1^p$&\cr
height4pt & \omit && \omit && \omit && \omit && \omit &\cr
&$D_{\ell>4,k>2}$&&$k,\ell\ {\rm even},\ {\rm and}\ k\ell \equiv_80$
&&$4$&&$\{C_1^a\si_{vsc}^b\}$&&${\Bbb D}_2$&\cr
height4pt & \omit && \omit && \omit && \omit && \omit &\cr
&$D_{\ell,k}$&&$\ell\ {\rm odd},\ k \equiv_4 0$&&$2$&&$\{\si_1,C_1\}$&&$
{\Bbb D}_1$&\cr
height4pt & \omit && \omit && \omit && \omit && \omit &\cr
&$D_{\ell,k>2}$&&$\ell\ {\rm odd},\ k \equiv_4 2$&&$4$&&$\{C_1^a\si_s^b\}$&&
${\Bbb D}_2$&\cr
height4pt & \omit && \omit && \omit && \omit && \omit &\cr
&$D_{\ell,k}$&&$k>1\ {\rm odd},\ \ell\not\equiv_84$&&$4$&&$\{C_1^a\si_v^b\}$&&
${\Bbb D}_2$&\cr
height4pt & \omit && \omit && \omit && \omit && \omit &\cr
&$D_{\ell,k>2}$&&$k\equiv_4\ell\equiv_4 2$&&
$8$&&$\{C_1^a\langle\si_s,\si_c\rangle\}$&&${\Bbb D}_4$&\cr
height4pt & \omit && \omit && \omit && \omit && \omit &\cr
&$D_{\l>4,k}$&&$k>1\ {\rm odd},\ \l\equiv_84$&&$12$
&&$\{C_1^a\langle\si_s,\si_c\rangle\}$&&${\Bbb D}_6$&\cr
height7pt & \omit && \omit && \omit && \omit && \omit &\cr}
\hrule}}}

\smallskip
\leftskip=1cm
\rightskip=1cm
\noindent
\baselineskip=12pt
{\bf Table 1.}  Complete list of automorphism invariants for classical simple
Lie algebras. The variables $c,p,t$ for $A_{\l,k}$ are defined in the text
(section 5). For $B_{\l,2}$, respectively $D_{\l,2}$, $p$ is the number of
distinct prime divisors of $2\l+1$, respectively $\l$. The exponents
$a,b$ range over $\{0,1\}$. We denote a congruence modulo $m$ by $\equiv _m$.
In the last column giving the structure of the automorphism group, we have
denoted by ${\Bbb D}_m$ the dihedral group of order $2m$.
\bigskip
\leftskip=0cm
\rightskip=0cm
\baselineskip=15pt
\endinsert

\topinsert
\centerline{
{\vbox{\offinterlineskip
\hrule
\halign{& \vrule# & \strut \quad \hfill#\hfill \quad
& \vrule# & \strut \quad \hfill#\hfill \quad
& \vrule# & \strut \quad \hfill#\hfill \quad
& \vrule# & \strut \quad \hfill#\hfill \quad
& \vrule# & \strut \quad \hfill#\hfill \quad & \vrule# \cr
height5pt & \omit && \omit && \omit && \omit && \omit &\cr
& $X_{\l,k}$ && conditions && \# autom. && names && group & \cr
height5pt & \omit && \omit && \omit && \omit && \omit &\cr
\noalign{\hrule}
height7pt & \omit && \omit && \omit && \omit && \omit &\cr
&$E_{6,k}$&&$k<3\ {\rm or}\ k\equiv_30$&&$2$&&$\{C^a\}$ &&${\Bbb D}_1$&\cr
height4pt & \omit && \omit && \omit && \omit && \omit &\cr
&$E_{6,k>2}$&&$k\equiv_3\pm 1$&&$4$&&$\{C^a\si_J^b\}$&&${\Bbb D}_2$&\cr
height7pt & \omit && \omit && \omit && \omit && \omit &\cr
\noalign{\hrule}
height7pt & \omit && \omit && \omit && \omit && \omit &\cr
&$E_{7,k}$&&$k=2\ {\rm or}\ k\not\equiv_4 2$&&$1$&&$\{\si_1\}$ &&&\cr
height4pt & \omit && \omit && \omit && \omit && \omit &\cr
&$E_{7,k}$&&$k>2\ {\rm and}\ k \equiv_4 2$&&$2$&&$\{\si_1,\si_J\}$&&${\Bbb
D}_1$&\cr
height7pt & \omit && \omit && \omit && \omit && \omit &\cr
\noalign{\hrule}
height7pt & \omit && \omit && \omit && \omit && \omit &\cr
&$E_{8,4}$&&&&$2$&&$\{\si_1,\si_{e8}\}$&&${\Bbb D}_1$&\cr
height4pt & \omit && \omit && \omit && \omit && \omit &\cr
&$E_{8,k}$&&$k\ne 4$&&$1$&&$\{\si_1\}$&&&\cr
height7pt & \omit && \omit && \omit && \omit && \omit &\cr
\noalign{\hrule}
height7pt & \omit && \omit && \omit && \omit && \omit &\cr
&$F_{4,3}$&&&&$2$&&$\{\si_1,\si_{f4}\}$&&${\Bbb D}_1$&\cr
height4pt & \omit && \omit && \omit && \omit && \omit &\cr
&$F_{4,k}$&&$k\ne 3$&&$1$&&$\{\si_1\}$&&&\cr
height7pt & \omit && \omit && \omit && \omit && \omit &\cr
\noalign{\hrule}
height7pt & \omit && \omit && \omit && \omit && \omit &\cr
&$G_{2,4}$&&&&$2$&&$\{\si_1,\si_{g2}\}$&&${\Bbb D}_1$&\cr
height4pt & \omit && \omit && \omit && \omit && \omit &\cr
&$G_{2,k}$&&$k\ne 4$&&$1$&&$\{\si_1\}$&&&\cr
height7pt & \omit && \omit && \omit && \omit && \omit &\cr}
\hrule}}}

\smallskip
\leftskip=1cm
\rightskip=1cm
\noindent
\baselineskip=12pt
{\bf Table 2.}  Complete list of automorphism invariants for exceptional simple
Lie algebras. The exponents $a,b$ range over $\{0,1\}$. A congruence modulo $m$
is denoted by $\equiv _m$. The notation ${\Bbb D}_m$ stands for the dihedral
group of order $2m$.
\bigskip
\leftskip=0cm
\rightskip=0cm
\baselineskip=15pt
\endinsert

Although scattered in the literature, all simple current automorphism
invariants have
been known for some time [3,1,13,30]. The exceptional automorphism invariant
of
$E_{8,4}$ was first given in [10], while those of $F_{4,3}$ and $G_{2,4}$ were
found in [35]. Finally all the exceptional automorphism invariants of
$B_{\l,2}$ and $D_{\l,2}$ have been recently unveiled in [12], though no
explicit description was given. Let us stress that all but one (namely the
$E_{8,4}$ exceptional) of the automorphism invariants of
Tables 1 and 2 can be fully accounted for in terms of simple currents,
conjugations and
Galois transformations. This is somewhat fortunate as they are the main
systematic procedures to construct automorphism invariants.

\smallskip
Our proof of this theorem relies on three basic steps. See the following
section for terminology.

\smallskip
We first examine the quantum dimensions $\D (\la) := S_{0,\la}/S_{0,0}$
for all weights in the alc\^ove. Let $[\la]$ denote the set of all transforms
--- the orbit --- of $\la$ by the symmetries of the extended Dynkin diagram;
$\D(\la)$ is constant along $[\la]$. It is well known that, as a function of
$\la$, $\D(\la)$ takes its minimal value 1 if and only if $\la\in[0]$ (for
$E_{8,2}$ there is an additional such $\la$, hence an additional simple
current,
but it plays no role here and will be ignored). Thus $\Q_1 = [0]$ is the
set of weights at which $\D(\la)$ is minimum. The first step of our proof is
to look for the set $\Q_2$ of all weights at which $\D(\la)$ takes its second
smallest value. In the generic case, we find that $\Q_2 = [\om^f]$ for
the fundamental weight $\om^f$ of $X_\ell$ which has the smallest
Weyl--dimension, in agreement with the large $k$ limit of the quantum
dimensions. If
however the level $k$ is sufficiently small, this simple statement may break
down, as Table 3 shows --- a prime example of that is given by the orthogonal
algebras at level 2. In these cases however, the spurious possibilities can
be handled by the norm condition (3.8a) and/or by looking at the sets
$\Q_i$ for $i \geq 3$, except for $B_{\l,2}$ and $D_{\l,2}$, which require a
special analysis. We refer the reader to the text for the details of these
cases. When $\Q_2 = [\om^f]$, we obtain our first conclusion that any
automorphism $\si$ must map the orbit $[\om^f]$ onto itself.

\smallskip
If $\Q_2 = [\om^f]$, we obtain from the first step that the action of $\si$ on
$\om^f$ is of the form $\si(\om^f) = C' J(\om^f)$ for some conjugation $C'$
and simple current $J$. A conjugation $C'$ always defines an automorphism
invariant, so that replacing $\si$ by $C' \circ \si$ permits us to assume
$\si(\om^f)=J(\om^f)$. Requiring that $\si$ commute with the modular matrix
$T$ --- the norm of the weights must be preserved --- puts various restrictions
on $J$, depending on the level $k$ and the algebra we consider.
Two situations are then possible. The first is that, for a given simple
current $J$ satisfying the norm condition, there does exist a simple current
automorphism invariant
$\si'$ such that $\si'(\om^f) = J(\om^f)$. In this case, the action of $J$ on
$\om^f$ lifts to an acceptable solution $\si'$ on the whole of the alc\^ove.
This means one may replace $\si$ by $\si'{}^{-1} \circ \si$, and assume that
$\si$ fixes $\om^f$. The second situation is when $J$ does not lift to a simple
current automorphism invariant.

\smallskip
The third and final step aims at filling the gaps left by the second step.
On the one
hand, we classify the automorphisms which leave $\om^f$ fixed. When combined
with the automorphisms  which do not leave $\om^f$ fixed --- these
were collected at Step 2 ---, they yield the full set of automorphisms. On the
other hand, we show that the possibilities $\si(\om^f) = J(\om^f)$ in the
second situation in Step 2 cannot be extended globally to any automorphism
invariant. The main tool to obtain these two
results is the explicit computation of fusion products. Indeed a happy feature
of $\om^f$ is that it is sufficiently small and simple to allow the calculation
of its fusion product with any other field, and this is what is basically
needed though in some cases the fusion with other small representations is also
required.

\smallskip
The first and crucial step of our proof is detailed in section 4, while the
other two are worked out in section 5 for the classical algebras, and in
section 7 for the exceptional ones. A section 6 is inserted that contains the
relevant analysis for $B_{\l,2}$ and $D_{\l,2}$. For completeness, we include
the results (but not the proofs) for the $A_\l$ series [17].

\vskip 1truecm
\noindent{\bf 3. Notations and Preliminaries}

\vskip 0.7truecm \noindent
Let $X_\l$ be a finite--dimensional simple Lie algebra. The weights are
denoted in
the Dynkin basis by $\la=(\la_1,\la_2,\ldots,\la_\l):=\sum_i\la_i\om^i$ with
all $\la_i$ integers, where $\om^i$ is the $i$--th fundamental weight. (Our
convention for the numbering of the simple roots is Dynkin's, as
used in [22].) The Weyl vector is $\rho=(1,1,\ldots,1)$. The
colabels $a_i^\vee$ are defined through the expansion of the highest root
$\psi$ in the basis of simple roots, $\psi = \sum_i ({2a_i^\vee \over
\alpha_i^2}) \alpha_i$. Put $a_0^\vee=1$. The dual Coxeter number $h^\vee =
1+\rho \cdot \psi = \sum_i a_i^\vee$.

By $X_{\l,k}$ we will mean the current algebra based on
$X_\l$, at a level $k \in {\Bbb N}$. The
height is defined by $n=k + h^\vee$. The integrable highest weight
representations of
$X_{\l,k}$ are in one--to--one correspondence with the set of dominant weights
 (also called the alc\^ove) of $X_{\l,k}$ given by [22]
$$
P_+(X_{\l,k}) = \big\{ \la=(\la_0;\la_1,\la_2,\ldots,\la_\l) \;|\; \la_i
\in {\Bbb N} \hbox{ and } \sum_{i=0}^\ell a_i^\vee \la_i = k \big\},
\eqno(3.1)
$$
and have characters denoted by $\chi_\la(\tau,z,u)$. For fixed level $k$, the
zero--th Dynkin label is redundant, so that two notations
$(\la_0;\la_1,\la_2,\ldots,\la_\l)$ and $(\la_1,\la_2,\ldots,\la_\l)$
designate a single element of $P_+(X_{\l,k})$. The identity 0 corresponds to
$k\om^0 :=(k;0,\ldots,0)$.

\smallskip
The characters
$\{\chi_\la\}_{\la \in P_+(X_{\ell,k})}$ transform linearly under the
action of $SL(2,{\Bbb Z})$, defined as follows by its generators
$(\tau,z,u) \mapsto (\tau+1,z,u)$ and $(\tau,z,u) \mapsto
({-1 \over \tau},{z \over \tau},u+{z^2\over 2\tau})$ [23]; these representing
matrices (called Kac-Peterson matrices) are respectively
$$\eqalignno{
& T_{\la,\la'} = \gamma \exp{\left( {2\pi i (\rho+\la)^2 \over 2n} \right)} \,
\delta_{\la,\la'}, &(3.2a) \cr
& S_{\la,\la'} = \gamma' \sum_{w \in W} \,({\rm det}\,w) \exp{ \left(
-{2\pi i (\rho +\la) \cdot w(\rho + \la') \over n} \right)}. &(3.2b) \cr}
$$
$\gamma$ and $\gamma'$ are constants independent of $\la$ and $\la'$, and $W$
is
the Weyl group of $X_\l$. The matrices $S$ and $T$ are both symmetric and
unitary, and satisfy $S^2=(ST)^3=C$. $C$, called the charge conjugation,
is an order 2 symmetry of the Dynkin diagram of $X_\l$ (if
non--trivial).

In particular the matrix elements $S_{0,\la}$ are all real and strictly
positive, and obey $S_{0,\la} \geq S_{0,0}$ for all $\la
\in P_+(X_{\l,k})$. The quantum dimension $\D(\la)$ is defined by
$$
\D(\la) :={S_{0,\la} \over S_{0,0}} = \prod_{\alpha >0} \;
{\sin{[\pi \alpha \cdot (\rho +\la)/n]} \over \sin{[\pi \alpha \cdot \rho
/n]}},
\eqno(3.3)
$$
where the product is over the positive roots of $X_\l$, and the second equality
follows from the Weyl denominator formula. Note that $\D(\la)
\geq 1$. Those weights $\la$ which satisfy $\D(\la)
= 1$ are called {\it simple currents} [31]. Except for one single case, namely
$E_{8,2}$, they are all given [9] by the action $\la = J(k\om^0)$ of
a symmetry $J$ of the extended Dynkin diagram which does not fix the
zero--th node; these $J$ act on weights by permuting their Dynkin labels.
By abuse of language, the same notation $J$ is used to denote
the simple current and the corresponding permutation. The simple currents
permute the weights in the alc\^ove and have the important property that
$$
S_{\la,\la'} = \exp{\big[ -2\pi iQ_J(\la')\big]} \, S_{J\la,\la'} =
\exp{\big[ -2\pi iQ_J(\la)\big]} \, S_{\la,J\la'},
\eqno(3.4)
$$
where the charge $Q_J(\la)$ and conformal weight $h_J$ are defined by
$$\eqalignno{
& Q_J(\la) \equiv h_\la + h_{J(0)} - h_{J(\la)}  \; \bmod 1, &(3.5a) \cr
& h_\la \equiv {(\rho + J(\la))^2 - \rho^2 \over 2n} \; \bmod 1. &(3.5b) \cr}
$$
The simple currents were classified in [9] for all simple Lie algebras.
Their explicit form will be given in the text (see sections 5 and 7).

\medskip
The weights in $P_+(X_{\l,k})$ form a ring, called the
fusion ring: the elements are {\it formal} linear combinations over ${\Bbb
Z}$ of the weights, and
the product $\la\times\mu=\sum N_{\la,\mu}^\nu\nu$ has non--negative
integer structure constants $N_{\la,\mu}^\nu$ called fusion coefficients.
These are defined by the Verlinde formula [34]
$$
N_{\la,\mu}^\nu = \sum_{\beta \in P_+(X_{\l,k})} \,
{S_{\la,\beta} S_{\mu,\beta} S^*_{\nu,\beta} \over S_{0,\beta}}.
\eqno(3.6a)
$$
They can be computed, at least in principle, by Lie algebraic
methods. For example [22,36,14],
$$
\eqalignno{N_{\la,\mu}^\nu=&\sum_{w\in \widehat W} ({\rm det}\ w) \,
R_{\la,\mu}^{w.\nu}&(3.6b)\cr
=&\sum_{\be\in P(\mu)}\sum_{{w\in\widehat W \atop
w.(\la+\be)=\nu}}({\rm det}\ w)\, {\rm mult}_\mu(\be),
&(3.6c)\cr}
$$
where $w.\nu=w(\nu+\rho)-\rho,$ $\widehat W$ is the (affine) Weyl group of
$X_{\l,k}$, and $R_{\la,\mu}^\nu \in {\Bbb N}$ are the Clebsch--Gordan series
coefficients of the $X_\l$ tensor product $\la \otimes \mu$. In (3.6c),
$P(\mu)$ is the set of weights of the $X_\l$ representation $\mu$, and
mult${}_\mu(\be)$ is the multiplicity of $\be$ in $\mu$. It should always
be clear from the context whether ``$\la+\mu$'' refers to the formal sum of
the fusion ring, or the usual component--wise sum.

\bigskip
In this paper, we will classify all modular invariant partition functions
$$
Z = \sum_{\mu,\mu' \in P_+(X_{\l,k})} \, M_{\mu,\mu'} \chi^*_\mu \,
\chi_{\mu'}
\eqno(3.7)
$$
for which the integer matrix $M$  defines
a permutation $\si$ of the alc\^ove by $M_{\mu,\mu'} = \delta_{\mu',\si
(\mu)}$. Modular invariance of (3.7) is equivalent to the statement that
$\si$ commutes with the matrices $S$ and $T$, that is,
$$\eqalignno{
& T_{\la,\la'} = T_{\si(\la),\si(\la')}, &(3.8a) \cr
& S_{\la,\la'} = S_{\si(\la),\si(\la')}. &(3.8b) \cr}
$$
Any permutation $\si$ of $X_{\l,k}$ obeying (3.8a),(3.8b) is called an
automorphism invariant. Note that they form a group under composition.
Since the 0--th row of $S$ is the only positive one, (3.8b) implies $\si$
will fix the identity,
$$
\si(0)=0.
\eqno(3.8c)
$$
{}From (3.6a) and (3.8b), $\si$ is an automorphism of the fusion ring:
$$
N_{\si(\la),\si(\mu)}^{\si(\nu)} = N_{\la,\mu}^\nu
\eqno(3.8d)
$$
(the converse is not true though). For this reason, the corresponding partition
functions are called permutation invariants or automorphism invariants.

We will denote the trivial permutation by $\si_1$. At present three main
methods of
systematically constructing non--trivial automorphism invariants are known:
conjugations, simple currents and Galois transformations can be used.
(In principle, these constructions are independent, but they
sometimes overlap, as has recently been discussed [12].) Any symmetry of
the Dynkin diagram which fixes the zero--th node is called a conjugation;
they act on weights by permuting their Dynkin labels, and as such always
define automorphism invariants.

\medskip
Simple currents provide a large stock of automorphism
invariants [31,20]. Let $N$ be the order of a simple current $J$. When $N
h_{J(0)}$ is an integer coprime with $N$, we can define a
simple current automorphism invariant by setting [32]
$$
\si_J(\la) = J^a(\la), \qquad \hbox{ with $ah_{J(0)} \equiv Q_J(\la) \bmod 1$
.} \eqno(3.9)
$$
It can be checked that $\si_J$ indeed commutes with $T$ and $S$, and is a
permutation of the alc\^ove.

Incidentally, many special cases of (3.9) were written down first by
[3,1,13]. Let us also mention that, when two
independent simple currents exist, a different
kind of simple current automorphism than (3.9) sometimes exists, called
an integer spin simple
current automorphism [30]. For simple $X_\ell$, this kind
of automorphism only
exists for the $D_\l$ series (it is denoted by $\si_{vsc}$ in Table 1), so that
we refrain from giving the general description and merely refer to the
$D_\l$--subsection 5.4 for its precise definition.

By $[\la]$ we mean the orbit of $\la$ under all the conjugations $C_i$ and
simple currents $J_j$. These orbits play an important role in this paper.

\medskip
Another way to construct modular invariants is by Galois
transformations. We see from (3.2b) (in fact this holds for any RCFT [7])
that the matrix elements
$S_{\la,\la'}$ lie in a cyclotomic extension of the rationals ${\Bbb
Q}(\zeta_N)={\Bbb Q}(\exp{2\pi i/N})$, for some algebra-dependent integer $N$.
Its Galois group is isomorphic to ${\rm Gal}({\Bbb Q}(\zeta_N)/{\Bbb Q}) \cong
\Z^*_N$, the group of invertible integers modulo
$N$. It is immediate from (3.2b) that any element $g$ of the Galois group
induces a
permutation $\la \mapsto g(\la)$ of the alc\^ove through its action on $S$
$$
g(S_{\la,\la'}) = \epsilon_g(\la) S_{g(\la),\la'} = \epsilon_g(\la')
S_{\la,g(\la')},
\eqno(3.10)
$$
where $\epsilon_g(\la)=\pm 1$ is a sign that only depends on $g$ and $\la$. The
images $g(\la)$ and $g(\la')$, the Galois transforms of $\la$ and $\la'$, can
be quite explicitly computed in the following way. Let $g_a$ with $a \in
\Z^*_N$
be a Galois transformation. Then $g_a(\la)$ is the unique weight in the
alc\^ove such that $\rho + g_a(\la) = w_{a,\la}\big(a(\rho + \la)\big) +
n\alpha^\vee_{a,\la}$ for some Weyl transformation $w_{a,\la}$ and some
$\alpha^\vee_{a,\la}\in Q^\vee$, the co-root lattice. Also
the sign appearing in (3.10) is given by $\epsilon_{g_a}(\la) = {\rm det}\,
w_{a,\la}$.

Under some conditions, Galois
transformations directly define automorphism invariants by setting
$M_{\la,\la'} =
\delta_{\la',g(\la)}$. This is the case whenever $g$ fixes the identity,
$g(0)=0$, and commutes with $T$ [11]. More generally, suppose that
$g$ is such that $g(0)=J(0)$ for some simple current $J$ and that $g$
commutes with $T$, and $g^2=1$. Then the following
defines an automorphism invariant:
$$
\si_g(\la)=\cases{J(g(\la)) & if $\epsilon_g(\la)=\epsilon_g(0)$, \cr
\noalign{\smallskip}
g(\la) & if $\epsilon_g(\la)=-\epsilon_g(0)$. \cr}
\eqno(3.11)
$$
The proof is simple. The extreme l.h.s. and r.h.s. of the equalities
$$
{\rm exp}[2\pi iQ_J(\la)]S_{0,\la} = S_{J(0),\la} = S_{g(0),\la} =
\epsilon_g(0) \, \epsilon_g(\la) \, S_{0,g(\la)},
\eqno(3.12)
$$
imply
$$
\exp[2\pi iQ_J(\la)] = \epsilon_g(0) \, \epsilon_g(\la).
\eqno(3.13)
$$
The same equation (3.12) with $\la$ replaced by $J(\la)$ shows that
$Q_J(J(\la)) \equiv Q_J(\la) \bmod 1$. Thus $Q_J(\la)$ can only take the values
0 and $1 \over 2$ modulo 1, and $J$ is a simple current of order 2, $J^2=id$.
Moreover acting with $g^2=1$ on $S_{0,\la}$, one obtains $Q_J(g(\la))
\equiv Q_J(\la) \bmod 1$. That $\si_g$ obeys (3.8a,b) can now be verified.

\smallskip
Equation (3.11) appears to be new. We will call the corresponding
$\si$ {\it generalized Galois automorphisms} since they reduce to pure
Galois automorphisms if $g(0)=0$. In section 6.3 we will show that the
$B_{\ell,2}$ and $D_{\ell,2}$ exceptional invariants have precisely this form.
Incidentally, in most cases (including all cases concerning us in this paper)
$[g,T]=0$ implies $g^2=1$. Moreover, the charge conjugation $C=S^2$ always
corresponds to $g_{-1}$.

\bigskip
We finish this section with a lemma which will be repeatedly used
throughout the paper. It is a slight generalization of a result proved in
[17].

\medskip \noindent
{\bf Lemma 1.} {\sl Let $\si$ be an automorphism invariant for $X_{\l,k}$.
If $\si$ fixes all $\om^i\in P_+(X_{\ell,k})$, then $\si$ is
the trivial permutation on $P_+(X_{\l,k})$.}

\bigskip \noindent
To prove this, it suffices to
show that any $S_{\la,\mu}/S_{0,\mu}$ can be written as a polynomial $P'_\la$
in the ratios $S_{\om^i,\mu}/S_{0,\mu}$ for all $\om^i \in P_+(X_{\ell,k})$.
This is true because from (3.8b) and the fact that the identity 0 and all
$\om^i$ are
fixed by $\si$, one obtains $S_{\la,\si(\mu)}=S_{\la,\mu}$ for all $\la,\mu$,
so that if $\si \neq 1$, two columns of $S$ would be equal and $S$ would be
singular. We do know from [17] that $S_{\la,\mu}/S_{0,\mu}$ can be
written as a polynomial $P_\la$ in the ratios $S_{\om^i,\mu}/S_{0,\mu}$ for all
$1\le i\le \ell$. The problem is that if $k$ is small, not all $\om^i$ may
lie in $P_+(X_{\ell,k})$ (we use (3.2b) to extend the definition of
$S_{\la,\mu}$ outside the alc\^ove).

Suppose $\nu\not\in
P_+(X_{\ell,k})$, for some weight $\nu$. Then either $\nu$ lies in a wall,
in which case
$$
S_{\nu,\mu}=0, \qquad \forall \mu\in P_+(X_{\ell,k}),
\eqno(3.14a)
$$
or there exists a $\beta_\nu \in P_+(X_{\ell,k})$, a $w_\nu\in W$, and an
element $\al_\nu^\vee$ of the co--root lattice such that
$\rho+\nu=w_\nu(\rho+\be_\nu)+n\al_\nu^\vee$, in which case
$$
S_{\nu,\mu} = ({\rm det}\,w_\nu)\,S_{\be_\nu,\mu}, \qquad \forall \mu\in P_+
(X_{\ell,k}).
\eqno(3.14b)
$$
All that we need to verify is that whenever $a^\vee_i>k$, either $\nu=\om^i$
satisfies (3.14a), or it satisfies (3.14b) with $\be_\nu=0$ or $\om^j$ for
some $j$. This is automatic whenever $a^\vee_i=k+1$, or when $P_+(X_{\ell,k})$
contains only weights of the form 0 and $\om^j$.

This leaves only $E_{7,2}$ with $\nu=\om^3$, and $E_{8,4}$ with $\nu=\om^5$. It
suffices to show that $\be_\nu\ne 2\om^6$, respectively $\om^1+\om^7$, $2\om^1$
or $2\om^7$. But $\rho+\nu$ and $\rho+\be_\nu$ must have the same norm modulo
$2n$, and checking the norms, we find that $\be_\nu$ cannot take these values,
so that Lemma 1 is proved for all $k$.

\eject
\vskip 2truecm
\noindent{\bf 4. Quantum Dimensions}

\vskip 0.7truecm \noindent
In this section we use quantum dimensions to find a weight $\om^f$ at
each level which must be fixed (up to extended Dynkin diagram symmetries)
by any automorphism invariant $\si$.

\smallskip
Recall the definition of quantum dimension $\D(\la)$, given in (3.3).
The positive roots $\alpha$ are explicitly given in e.g. [4].
Let ${\cal Q}_1$ be the set of all weights $\la\in P_{+}(X_{\ell,k})$ with
the smallest value of ${\cal D}(\la)$, let ${\cal Q}_2$
be those with the second smallest value, etc. We know that for all
$\la'\in[\la]$, $\D(\la)=\D(\la')$.

By (3.8b),(3.8c), we find that $\D(\la)=
\D(\si\la)$, hence
$$
\si{\cal Q}_m={\cal Q}_m,\qquad \forall m=1,2,\ldots
\eqno(4.1a)
$$
Fuchs [9] found the set ${\cal Q}_1$ for any $X_{\ell,k}$: in all cases
except one, ${\cal Q}_1=[0]$; the only exception is $E_{8,2}$, where
${\cal Q}_1=[0]\cup[\om^7]$. He proved this by regarding $\D(\la)$
as an analytic function of $\ell$ real variables $\la_1,\ldots,\la_\ell$,
defined
by the expression in (3.3). These $\la$ are to lie in the convex hull
$$
\overline{P}_{+}(X_{\ell,k}) := \bigl\{\sum_{i=0}^\ell\la_i\om^i \;|\;
\la_i \in {\Bbb R}_{\ge}\hbox{ and }\sum_{i=0}^\ell \la_i a^\vee_i = k \bigr\}.
$$
It was found in [9] that, for all $i=1,\ldots,\ell$,
$$
{\partial\over\partial \la_i}\D(\la)=0 \qquad \Longrightarrow \qquad
{\partial^2\over \partial \la_i\partial\la_j} \D(\la)<0, \qquad
\forall j=1,\ldots,\ell.
\eqno(4.1b)
$$
Though (4.1b) is not strong enough for our purposes, this basic idea
will be a critical step in our analysis.

\smallskip
The main result of this section is the determination of ${\cal Q}_2$ for
all $X_{\ell,k}$.

\bigskip \noindent
{\bf Proposition.} \quad {(a)} {\sl for $X_\ell=A_\ell$, $\ell \ge 1$ and
$k \ge
2$ \ :\ \  ${\cal Q}_2 =[\om^1]$,}

\leftskip 0.5cm
\item{(b)} {\sl for $X_\ell=B_\ell$, $\ell \ge 3$ and $k\ge 4$ \ :\ \
${\cal Q}_2 =[\om^1]$,}

\item{(c)} {\sl for $X_\ell=C_\ell$, $\ell \ge 2$ and $k=1$ or
$\l+k \ge 6$ \ :\ \   ${\cal Q}_2=[\om^1]$, }

\item{(d)} {\sl for $X_\ell=D_{\l \ge 4}$ and $E_6$, $k\ge 3$ \ :\ \
${\cal Q}_2=[\om^1]$,}

\item{(e)} {\sl for $X_\ell=E_7$, $E_8$, $F_4$ and $G_2$, $k \ge 5$ \ :\ \
${\cal Q}_2=[\om^6]$, $[\om^1]$, $[\om^4]$ and $[\om^2]$ respectively.}

\leftskip 0cm
\bigskip \noindent
For the levels missed by the Proposition, we have listed in Table 3 the sets
$\Q_m$ for small $m$.
Together with the $T$--condition (3.8a) and the selection rule (4.1a), the
Proposition and Table 3 give us the following valuable facts.

\topinsert
\centerline{
{\vbox{\offinterlineskip
\hrule
\halign{& \vrule# & \strut \quad \hfill#\hfill \quad
& \vrule# & \strut \quad \hfill#\hfill \quad
& \vrule# & \strut \quad \hfill#\hfill \quad
& \vrule# & \strut \quad \hfill#\hfill \quad
& \vrule# & \strut \quad \hfill#\hfill \quad
& \vrule# & \strut \quad \hfill#\hfill \quad
& \vrule# & \strut \quad \hfill#\hfill \quad & \vrule# \cr
height5pt & \omit && \omit && \omit && \omit && \omit && \omit && \omit &\cr
& $X_{\l}$ && $k$ && ${\cal Q}_1$ && ${\cal Q}_2$ && ${\cal Q}_3$ &&
${\cal Q}_4$ && ${\cal Q}_5$ &\cr
height5pt & \omit && \omit && \omit && \omit && \omit && \omit && \omit &\cr
\noalign{\hrule}
height5pt & \omit && \omit && \omit && \omit && \omit && \omit && \omit &\cr
& $A_\ell$ && 1 && $[0]=[\om^1]$ &&&&&&&&&\cr
height5pt & \omit && \omit && \omit && \omit && \omit && \omit && \omit &\cr
\noalign{\hrule}
height5pt & \omit && \omit && \omit && \omit && \omit && \omit && \omit &\cr
&$B_\ell$&& 1&&$[0]=[\om^1]$&&&&&&&&&\cr
height4pt & \omit && \omit && \omit && \omit && \omit && \omit && \omit &\cr
&&&2&&$[0]$&&$[\om^1]\cup\cdots\cup[\om^{\ell-1}]\cup[2\om^\ell]$&&&&&&&\cr
height4pt & \omit && \omit &&\omit && \omit && \omit && \omit && \omit &\cr
&&& 3&&$[0]$&&$[3\om^\ell]$&&$[\om^1]$&&&&&\cr
height5pt & \omit && \omit && \omit && \omit && \omit && \omit && \omit &\cr
\noalign{\hrule}
height5pt & \omit && \omit && \omit && \omit && \omit && \omit && \omit &\cr
&$C_2$&& 2&&$[0]$&&$[\om^2]\cup[2\om^1]$&&$[\om^1]$&&&&&\cr
height4pt & \omit && \omit && \omit && \omit && \omit && \omit && \omit &\cr
&$C_2$&& 3&&$[0]$&&$[\om^1]\cup[\om^2]\cup[3\om^1]$&&&&&&&\cr
height4pt & \omit && \omit && \omit && \omit && \omit && \omit && \omit &\cr
&$C_3$&& 2&&$[0]$&&$[\om^1]\cup[\om^3]\cup[2\om^1]$&&&&&&&\cr
height5pt & \omit && \omit && \omit && \omit && \omit && \omit && \omit &\cr
\noalign{\hrule}
height5pt & \omit && \omit && \omit && \omit && \omit && \omit && \omit &\cr
&$D_\ell$&& 1&&$[0]=[\om^1]$&&&&&&&&&\cr
height4pt & \omit && \omit && \omit && \omit && \omit && \omit && \omit &\cr
&$D_4$&& 2&&$[0]$&&$[\om^1]\cup\cdots\cup[\om^\ell]$&&&&&&&\cr
height4pt & \omit && \omit && \omit && \omit && \omit && \omit && \omit &\cr
&$D_{\l>4}$&& 2&&$[0]$&&$[\om^1]
\cup\cdots\cup[\om^{\ell-1}]$&&&&&&&\cr
height5pt & \omit && \omit && \omit && \omit && \omit && \omit && \omit &\cr
\noalign{\hrule}
height5pt & \omit && \omit && \omit && \omit && \omit && \omit && \omit &\cr
&$E_6$&& 1&&$[0]=[\om^1]$&&&&&&&&&\cr
height4pt & \omit && \omit && \omit && \omit && \omit && \omit && \omit &\cr
&&& 2&&$[0]$&&$[\om^2]$&&$[\om^1]$&&&&&\cr
height5pt & \omit && \omit && \omit && \omit && \omit && \omit && \omit &\cr
\noalign{\hrule}
height5pt & \omit && \omit && \omit && \omit && \omit && \omit && \omit &\cr
&$E_7$&& 1&&$[0]=[\om^6]$&&&&&&&&&\cr
height4pt & \omit && \omit && \omit && \omit && \omit && \omit && \omit &\cr
&&& 2&&$[0]$&&$[\om^7]$&&$[\om^1]$&&$[\om^6]$&&&\cr
height4pt & \omit && \omit && \omit && \omit && \omit && \omit && \omit &\cr
&&& 3&&$[0]$&&$[\om^1]\cup[\om^2]\cup[\om^6]$&&&&&&&\cr
height4pt & \omit && \omit && \omit && \omit && \omit && \omit && \omit &\cr
&&& 4&&$[0]$&&$[2\om^7]$&&$[\om^6]$&&&&&\cr
height5pt & \omit && \omit && \omit && \omit && \omit && \omit && \omit &\cr
\noalign{\hrule}
height5pt & \omit && \omit && \omit && \omit && \omit && \omit && \omit &\cr
&$E_8$&& 1&&$[0]$&&&&&&&&&\cr
height4pt & \omit && \omit && \omit && \omit && \omit && \omit && \omit &\cr
&&& 2&&$[0]\cup[\om^7]$&&$[\om^1]$&&&&&&&\cr
height4pt & \omit && \omit && \omit && \omit && \omit && \omit && \omit &\cr
&&& 3&&$[0]$&&$[\om^8]$&&$[\om^2]$&&$[\om^1]$&&&\cr
height4pt & \omit && \omit && \omit && \omit && \omit && \omit && \omit &\cr
&&& 4&&$[0]$&&$[2\om^7]$&&$[2\om^1]$&&$[\om^1]\cup[\om^6]$&&&\cr
height5pt & \omit && \omit && \omit && \omit && \omit && \omit && \omit &\cr
\noalign{\hrule}
height5pt & \omit && \omit && \omit && \omit && \omit && \omit && \omit &\cr
&$F_4$&& 1&&$[0]$&&$[\om^4]$&&&&&&&\cr
height4pt & \omit && \omit && \omit && \omit && \omit && \omit && \omit &\cr
&&& 2&&$[0]$&&$[\om^1]$&&$[2\om^4]$&&$[\om^3]$&&$[\om^4]$&\cr
height4pt & \omit && \omit && \omit && \omit && \omit && \omit && \omit &\cr
&&& 3&&$[0]$&&$[\om^1]\cup[3\om^4]$&&$[\om^2]\cup [\om^4]$&&&&&\cr
height4pt & \omit && \omit && \omit && \omit && \omit && \omit && \omit &\cr
&&& 4&&$[0]$&&$[\om^1]\cup[\om^4]\cup[2\om^1]\cup[4\om^4]$&&&&&&&\cr
height5pt & \omit && \omit && \omit && \omit && \omit && \omit && \omit &\cr
\noalign{\hrule}
height5pt & \omit && \omit && \omit && \omit && \omit && \omit && \omit &\cr
&$G_2$&& 1&&$[0]$&&$[\om^2]$ &&&&&&&\cr
height4pt & \omit && \omit && \omit && \omit && \omit && \omit && \omit &\cr
&&& 2&&$[0]$&&$[\om^1]$&&$[2\om^2]$&&$[\om^2]$&&&\cr
height4pt & \omit && \omit && \omit && \omit && \omit && \omit && \omit &\cr
&&& 3&&$[0]$&&$[\om^1]\cup[\om^2]\cup[3\om^2]$&&&&&&&\cr
height4pt & \omit && \omit && \omit && \omit && \omit && \omit && \omit &\cr
&&& 4&&$[0]$&&$[\om^2]\cup[2\om^1]$&&&&&&&\cr
height5pt & \omit && \omit && \omit && \omit && \omit && \omit && \omit &\cr}
\hrule}}}

\smallskip
\leftskip=1cm
\rightskip=1cm
\noindent
\baselineskip=12pt
{\bf Table 3.}  Quantum dimensions for small $k$. Here are listed those
exceptional cases missing in the Proposition, and the order on the
orbits of the weights, up to $[\om^f]$, induced by their quantum dimensions.
\bigskip
\leftskip=0cm
\rightskip=0cm
\baselineskip=15pt
\endinsert

\medskip \noindent
{\bf Corollary.} {\sl An automorphism invariant $\si$ necessarily satisfies:}

\leftskip 0.5cm
\item{(a)} {\sl $\si\om^1 \in [\om^1]$ for $A_\ell$, $C_\ell$ or $E_6$, any
$k$,}

\item{(b)} {\sl $\si\om^6 \in [\om^6]$ for $E_7$, any $k$,}

\item{(c)} {\sl $\si\om^1 \in [\om^1]$ for $B_\ell$ and $D_\ell$, any
$k \ne 2$,}

\item{(d)} {\sl $\si\om^4=\om^4$ for $F_4$, any $k \ne 3$,}

\item{(e)} {\sl $\si\om^1=\om^1$ and $\si\om^2=\om^2$ for $E_8$ and $G_2$
respectively, any $k \ne 4$.}

\leftskip 0cm
\bigskip
In what follows, we will denote by $\om^f$ the weight singled out
by the Proposition and Corollary --- so $\omega^f=\om^1$ for all but
$E_7,F_4,G_2$.
Note that in all cases $\om^f$ is the weight of $X_\ell$ with second smallest
Weyl dimension. This is of course not a coincidence, and happens because, for
fixed $\la$, ${\rm lim}_{k\rightarrow \infty} \D(\la)$ is the Weyl dimension
of $\la$.

The remainder of this section is devoted to the proof of the Proposition.

\bigskip\noindent
{\it Step 1}. The first step in the proof of the Proposition
will be to analyse (3.3), in order to come up with
a small list of candidates $\la\in P_{+}(X_{\ell,k})$ for belonging to
${\cal Q}_2$.

\smallskip
Choose any constants $a=\sum_{i=0}^\ell a_i\om^i$, $b=\sum_{i=0}^\ell b_i\om^i
\neq 0$, $a_i,b_i \in {\Bbb R}$. Suppose that for all $t \in [t_0,t_1]$,
$a+bt \in \overline{P}_{+} (X_{\ell,k})$. Then for $t_0\le t'\le t_1$, an easy
calculation gives
$$\eqalignno{
{d\over dt}\D(a+bt)\bigl|_{t=t'}=0 & \qquad \Longrightarrow & (4.2a)\cr
&{d^2\over dt^2}\D(a+bt)\bigr|_{t=t'}=-\D(a+bt')\,{\pi^2\over n^2}\,
\sum_{{\al}>0}{(b\cdot\al)^2\over \sin^2[\pi\,(a+bt'+\rho)\cdot\al/n]}<0.&\cr}
$$
This means that $\D(a+b\,t)$ will attain its minimum at one of the
endpoints $t=t_0,t_1$:
$$
{\rm for\ all}\ t_0<t'<t_1,\quad \D(a+b\,t')>{\rm min}\{\D(a+b\,t_0),\,
\D(a+bt_1)\}.
\eqno(4.2b)
$$
This implies the following rule. Suppose
$$
\sum_{i=0}^\ell m_ia_i^\vee=0, \qquad m_i \in {\Bbb Z},
\eqno(4.2c)
$$
and not all $m_i=0$. If $\la\in{\cal Q}_2$ and both $\la\pm m\not\in\Q_1$, then
$$
\la_i<|m_i| \quad {\rm for\ some}\ 0\le i\le \ell.
\eqno(4.2d)
$$
To prove this, we take a weight $\la$ of $P_+(X_{\l,k})$ and consider the
family $\la(t)=\la+mt$. With $t_0={\rm max}_{i\,:\,m_i>0}(-\la_i/m_i)$ and
$t_1={\rm min}_{i\,:\,m_i<0}(-\la_i/m_i)$, the weights $\la(t)$ belong to
$\overline{P}_{+} (X_{\ell,k})$ for all $t \in [t_0,t_1]$. Note that
$\la(t)$ will belong to $P_+(X_{\ell,k})$ if $t\in[t_0,t_1]$ is integer. If
both $\pm 1 \in [t_0,t_1]$, one obtains from (4.2b) a contradiction to $\la
\in {\cal Q}_2$ unless $\la(\pm 1) \in {\cal Q}_1$. Thus unless one of $\la
\pm m \in {\cal Q}_1$, we must have $t_0 > -1$ or $t_1 < 1$, implying (4.2d).

Writing $a_{ij}={\rm gcd}(a_i^\vee,a_j^\vee)$, a special case of (4.2d)
is that, for each choice of $0\le i<j \le \ell$,
$$
{\rm either}\quad \la_i<{a_j^\vee\over a_{ij}} \quad {\rm or} \quad
\la_j< {a^\vee_i\over a_{ij}},
\eqno(4.2e)
$$
again provided $\la\pm({a^\vee_j\over a_{ij}}\om^i-{a_i^\vee\over a_{ij}}\om^j
)\not\in\Q_1$. Eq.(4.2e) implies that if $\la\in
{\cal Q}_2$, then at most one $\la_i$ can be larger than ${\rm max}_j \,
a_j^\vee$.

\smallskip
Any $\la$ which obeys (4.2d) for all choices of $m_i$ satisfying
eq.(4.2c), will be called a {\it candidate}. Step 1 consists of finding
all candidates. The result is given in the lemma below, where we use the
following notation. Define
the {\it truncation} $[c]$ to be the largest integer not greater than
$c$, and the {\it remainder} $\{c\}_d$ to be $c-d[c/d]$. By $\mu(ij)$
we mean the weight
$$
\mu(ij): = \{k\}_{a_{ij}}\om^0 + x\om^i + y\om^j,
\eqno(4.3a)
$$
where $x$ and $y$ are given by
$$
x=\bigl\{[k/a_{ij}](a_i')^{-1}\bigr\}_{a_j'}\;,
\qquad y={k - \{k\}_{a_{ij}} - a_i^\vee x \over a^\vee_j}.
\eqno(4.3b)
$$
In (4.3b), $a_i' = a_i^\vee/a_{ij}$ and $a_j' = a^\vee_j/a_{ij}$,
and by $(a_i')^{-1}$ we mean the (integer) multiplicative
inverse mod $a_j'$. For example, if $a_i^\vee=2$ and $a^\vee_j=3$, we
get $(x,y)=(0,{k\over 3})$, $(2,{k-4 \over 3})$, $(1,{k-2 \over 3})$ for $k
\equiv 0,1,2 \bmod 3$, respectively, while if $a_i^\vee=3$ and $a_j^\vee=2$ we
get $(x,y)= (0,{k\over 2})$, $(1,{k-3\over 2})$ for $k \equiv 0,1 \bmod 2$,
respectively. Note that $\mu(ij)=\mu(0j)$ if $a_j^\vee$ divides either
$a_i^\vee$ or $k$.

The virtue of (4.3a) is that it gives in one formula almost all candidates
which have at most three non--zero Dynkin labels, one of them being $\la_0$ if
it has exactly three non--zero labels. Suppose for instance $\la_0,\la_i,\la_j
\neq 0$. Then from (4.2d), the choices $(m_0,m_i)=(a_i^\vee,-1)$,
$(m_0,m_j)=(a_j^\vee,-1)$ and $(m_0,m_i,m_j)=(a_i^\vee-a_j^\vee,-1,1)$ (all
others zero in each case) lead to $\la_0 < {\rm min}\{a_i^\vee,a_j^\vee,
|a_i^\vee-a_j^\vee|\}$. Checking all possible pairs of colabels, one can see
that it implies
$\la_0 < a_{ij}$, except for $E_8$ if $\{a_i^\vee,a_j^\vee\}=\{2,5\}$ or
$\{3,5\}$. Moreover
(4.2e) implies $\la_i < a_j'$ or $\la_j < a_i'$, say $\la_i < a_j'$ for
definiteness. Then
$$
\la_0 + a_i^\vee\la_i + a_j^\vee\la_j = k \quad \Longrightarrow \quad
\cases{\la_0 \equiv k \bmod a_{ij}, & \cr \noalign{\smallskip}
\la_i \equiv {k-\la_0 \over a_{ij}} (a_i')^{-1} \bmod a_j'. & \cr}
\eqno(4.3c)
$$
When $\la_0<a_{ij}$, the r.h.s. of (4.3c) uniquely fixes $\la_0$ and $\la_i$,
which then determines the value of $\la_j$ using the l.h.s. of (4.3c) --
that is, in this case we find that indeed $\la=\mu(ij)$. Finally for $E_{8,k}$,
$k\equiv 3,4 \bmod 5$, there are four candidates with $\la_0 \ge a_{ij}$,
given separately in Lemma 2.

\medskip \noindent
{\bf Lemma 2.} \quad (1) {\sl The candidates for $A_{\ell,k}$ are
$[\om^i]$, $1\le i\le {\ell+1\over 2}$.}

\leftskip 0.5cm
\item{(2)} {\sl The candidates for $B_{\ell,k}$ are \ :\ \  $[\om^i]$ for
$1\le i\le
\ell$, $[\mu(0j)]$ and $[\mu(\ell j)]$ for $1<j<\ell$, and $[\mu(0\ell)]$.}

\item{(3)} {\sl The candidates for $C_{\ell,k}$ are \ :\ \  $[\om^i]$ for
$1\le i\le \ell$ and $[\mu(0j)]$ for $1\le j\le \ell/2$.}

\item{(4)} {\sl The candidates for $D_{\ell,k}$ are \ :\ \  $[\om^i]$ for
$1\le i< \ell$ and $[\mu(0j)]$ for $1<j<\ell-1$.}

\item{(5)} {\sl The candidates for $E_{6,k}$ are \ :\ \  $[\om^i]$ for
$i=1,2,3,6$, $[\mu(0j)]$ for $j=2,3,6$, $[\mu(23)]$ and $[\mu(32)]$.}

\item{(6)} {\sl The candidates for $E_{7,k}$ are \ :\ \  $[\om^i]$ for $1\le
i\le 7$, $[\mu(ij)]$ for most pairs $0\le i \neq j\le 7$, and
$[\om^{1,7}+\om^{2,4}+{k-5\over 4}\om^3]$ for $k \equiv 1 \bmod 4$.}

\item{(7)} {\sl The candidates for $E_{8,k}$ are \ :\ \  $[\om^i]$ for
$1\le i\le 8$, $[\mu(ij)]$ for most pairs $0\le i \neq j\le 8$, as well
as }

\item{} {\sl for $k\equiv 1 \bmod 4$, \quad $\om^{1,7}+\om^{2,8}+{k-5\over 4}
\om^{3,6}$ \quad and \quad $\om^{2,8}+{k-9\over 4}\om^{3,6}+\om^5$,

\item{} for $k\equiv 3 \bmod 4$, \quad $\om^{1,7}+{k-7\over 4}\om^{3,6}+
\om^4$ \quad and \quad ${k-11\over 4}\om^{3,6}+\om^4+\om^5$,

\item{} for $k\equiv 1 \bmod 5$, \quad $\om^{1,7}+\om^{3,6}+{k-6\over 5}\om^4$,

\item{} for $k\equiv 2 \bmod 5$, \quad $\om^{2,8}+\om^{3,6}+{k-7\over 5}\om^4$,

\item{} for $k\equiv 3 \bmod 5$, \quad $\om^{1,7}+{k-3\over 5}\om^4$,\quad
and \quad $\om^{1,7}+{k-8\over 5}\om^4+\om^5$,

\item{} for $k\equiv 4 \bmod 5$, \quad $\om^{2,8}+{k-4\over 5}\om^4$
\quad and \quad $\om^{2,8}+{k-9\over 5}\om^4+\om^5$,

\item{} for $k\equiv 1,2 \bmod 6$, \quad
$\om^{2,8}+(3-\{k\}_6)\om^4+[{k-8\over 6}]
\om^5$,

\item{} for $k\equiv 1,3 \bmod 6$, \quad ${1+\{k\}_6\over 2}\om^{1,7}+\om^4+
[{k-6\over 6}]\om^5$ \quad and \quad ${5-\{k\}_6\over 2}\om^{3,6}+\om^4+
[{k-9\over 6}]\om^5$,

\item{} for $k\equiv 1,5 \bmod 6$, \quad ${9-\{k\}_6\over 4}\om^{1,7}+\om^{2,8}
+[{k-5\over 6}]\om^5$ \quad and \quad $\om^{2,8}+{3+\{k\}_6\over 4}\om^{3,6}+
[{k-6\over 6}]\om^5$,

\item{} for $k=15$, \quad $2\om^{2,8}+\om^{3,6}+\om^4$.}

\item{(8)} {\sl The candidates for $F_{4,k}$ are \ :\ \  $\om^i$ for
$i=1,2,3,4$,
$\mu(0j)$ for $j=1,2,3,4$, $\mu(4j)$ for $j=1,2,3$, and
$\mu(12)$, $\mu(21)$, $\mu(23)$, $\mu(32)$. }

\item{(9)} {\sl The candidates for $G_{2,k}$ are \ :\ \  $\om^1$, $\om^2$,
$\mu(01)$, $\mu(02)$ and $\mu(21)$.}

\leftskip 0cm
\medskip
We use the notation $\om^{1,7}$, etc, to denote {\it either}
$\om^1$ {\it or} $\om^7$ (but not both simultaneously). Lemma 2 holds for
any $k$, though for small $k$ not all of these weights
will lie in $P_+(X_{\ell,k})$. Also, for some $k$ these candidates will
not all be distinct: e.g.\ for $B_{\ell,k}$, $k$ even, $\mu(\ell j)=\mu(0j)$.

\medskip
We will sketch the proof for the hardest case, namely $E_{8}$. First note
that by (4.2d), at most one element in each of the pairs $(\la_1,\la_7)$,
$(\la_2,\la_8)$, $(\la_3,\la_6)$ can be different from zero. For notational
convenience, suppose that $\la_6=\la_7=\la_8=0$.

Together with (4.2d), the seven arithmetic identities
$$
0=1+2-3=1+3-4=2+3-5=1+5-6=1+4-5=2+4-6=3-4-5+6
$$
tell us that at most three among $\la_0,\ldots,\la_5$ can be non--zero. If only
one or two of the $\la_i$, for $i>0$, are non--zero, then $\la$ will equal
either $\om^i$ or $\mu(ij)$, or equal $\om^1+{k-3\over 5}\om^4$ or
$\om^2+{k-4\over 5}\om^4$, as we have seen. Thus we may assume here that
exactly three of $\la_i$ are non--zero, and $\la_0=\la_6=\la_7=\la_8=0$.

Now we just run through the various possibilities. For example, suppose
$\la_1,\la_2,\la_3\ne 0$. From (4.2e) we have $\la_1=1$. Since $-2+2 \cdot 3-4
=0$, the inequality (4.2d) requires $\la_2=1$. Then the fact that $\la$ should
be in the alc\^ove at all, fixes $\la_3$ and constrains $k$. For another
example, suppose $\la_3,\la_4,\la_5 \ne 0$. Then $-4+2 \cdot 5-6=0$, so
that (4.2d) forces $\la_4=1$ and (4.2e) requires either $\la_3\le 2$ or
$\la_5=1$.

\bigskip \noindent
{\it Step 2.} Here we will use rank--level duality
of the quantum dimensions [28] to significantly reduce the numbers of
candidates given in Lemma 2, for $A_\ell$, $B_\ell$, $C_\ell$, $D_\ell$.

There is a well--known duality between the quantum dimensions of $A_{\ell,k}$
and $A_{k-1,\ell+1}$, $C_{\ell,k}$ and $C_{k,\ell}$, and $SO(m)_k$ and
$SO(k)_m$. In particular, writing $X_{\ell,k}\leftrightarrow X'_{k',\ell'}$,
we have
$$2^a\,\D(\la)=\D'(\la'),\eqno(4.4)$$
where $\D'$ is the quantum dimension for the dual theory $X'_{k',\ell'}$.
In all cases, the weight $0$ for $X_{\ell,k}$ is sent to the weight $0$ for
$X'_{k',\ell'}$, and $(\la')'\in[\la]$.
For $X_\ell=A_\ell$ or $C_\ell$, $a=0$ and $\la'$ is defined by saying
its Young tableau is the transpose of that of $\la$ (for this purpose we
may identify $C_1$ with $A_1$). The situation for $X_\ell=B_\ell$ and $D_\ell$
is slightly more complicated; we will give below all relevant values of
$\la'$ and $a$.\smallskip

When $X_\ell=B_\ell$ and $k>6$, for each $1\le j<\ell$, $(\om^j)'=j\om'{}^1$
with $a=0$.
Also, for each $1<j<\ell$, $([k/2]\om^j)'=2j\om'{}^{k'}$ with $a=0$ for
$k$ odd and $a=-1$ for $k$ even. For $k$ even, $(k\om^\ell)'=2\ell\om'
{}^{k'}$
with $a=-1$, while for $k$ odd, $(k\om^\ell)'=\om'{}^{k'}$ with $a=-{1\over
2}$. Finally, when $k$ is odd, $(\om^\ell)'=(2\ell+1)\om'{}^{k'}$ with
$a={1\over 2}$, and for each $1<j<\ell$, $\mu(\ell j)'=(2\ell+1-2j)\om'
{}^{k'}$ with $a=-{1\over 2}$.

When $X_\ell=D_\ell$ and $k>6$, for each $1\le j<\ell-1$, $(\om^j)'=j\om'{}^1$
with $a=0$, and for $1<j<\ell-1$, $([k/2]\om^j)'=2j\om'{}^{k'}$ with $a=0$
if $k$ is odd, and $a=-1$ if $k$ is even.

This rank-level duality for $X_\ell=B_\ell$ and $D_\ell$ extends to
$3\le k\le 6$ provided: we identify $B_{1,m}$ with $A_{1,2m}$ and put
$\om'{}^1:=2\tilde{\om}^1$, $\om'{}^{k'}:=\tilde{\om}^1$;
we identify $D_{2,m}$ with $A_{1,m}\oplus A_{1,m}$ and put
$\om'{}^1:=\tilde{\om}^1+\tilde{\om}^2$, $\om'{}^{k'}:=\tilde{\om}^1$;
we identify $B_{2,m}$ with $C_{2,m}$ and put
$\om'{}^1:=\tilde{\om}^2$, $\om'{}^{k'}:=\tilde{\om}^1$;
and we identify $D_{3,m}$ with $A_{3,m}$ and put
$\om'{}^1:=\tilde{\om}^2$, $\om'{}^{k'}:=\tilde{\om}^1$.
By $\tilde{\om}^i$ here we mean the fundamental weights for $A_1$, $A_1\oplus
A_1$, $C_2$, and $A_3$, respectively.

\medskip   Now we turn to the consequences of this rank-level duality for
finding ${\cal Q}_2$.
Consider first $X_\l=C_\l$. Since the duality here between quantum
dimensions is exact ({\it i.e.} $a=0$ always),  we have
$\la\in{\cal Q}_2$ iff $\la'\in{\cal Q}'_2$. This gives us an additional
constraint on $\la\in{\cal Q}_2$: $\la'$ must be a candidate of $C_{k,\l}$.
However, $(\om^j)'=j\om'{}^1$ and $(k\om^j)'=j\om'{}^k$, so
of these only $\om^1$, $\om^\ell$ and $k\om^1$ are the duals of candidates.
$X_\l=A_\l$ is similar.

The argument for $X_\l=B_\l$ and $D_\l$ is not much more difficult.
Consider for example $B_\l$ when $k>1$ is odd, and any $1<j<\l$:
$$\eqalignno{
\D(\om^j)=&\D'(j\om'{}^1)>{\rm min}\{
\D'(\om'{}^1),\,\D'(2\ell\om'{}^1)\}=\D'(\om'{}^1)=\D(\om^1), &(4.5a)\cr
\D({k-1\over 2}\om^j)=&\D'(2j\om'{}^{k'})>{\rm min}\{\D'(\om'{}^{k'}),\,
\D'((2\l+1)\om'{}^{k'})\}&\cr
=&{\rm min}\{\D({k-1\over 2}\om^1),\,\sqrt{2}\D(\om^\ell)\}\ge{\rm min}
\{\D(\om^1),\,\D(\om^\l)\},&(4.5b)\cr
\D({k-1\over 2}\om^j+\om^\ell)&=\sqrt{2}\,
\D'((2\ell+1-2j)\,\om'{}^{k'}) &(4.5c)\cr
>&\sqrt{2}\,{\rm min}\{\D'((2\ell+1)\om'{}^{k'}),\,\D'
(\om'{}^{k'})\}\le{\rm min}\{\D(k\om^\ell),\,\D(\om^\ell)\}.&\cr}
$$
In deriving (4.5) we use both rank--level duality and (4.2b).

\medskip
Summarizing, we find the following results:

\leftskip 1.5cm
\item{(1)} for $A_{\ell,k}$ and $k\ge 2$ \ :\ \  $\Q_2=[\om^1]$,
\item{(2)} for $B_{\ell,k}$ and $k\ge 3$ \ :\ \  $\Q_2\subseteq[\om^1]\cup[
\om^\ell]\cup[k\om^\ell]$,
\item{(3)} for $C_{\ell,k}$ \ :\ \  $\Q_2\subseteq[\om^1]\cup[\om^\ell]\cup
[k\om^1]$,
\item{(4)} for $D_{\ell,k}$ and $k\ge 3$ \ :\ \  $\Q_2\subseteq[\om^1]\cup
[\om^\ell]$.

\leftskip 0cm

\bigskip \noindent
{\it Step 3.} The remaining candidates $\la$ come in two forms.
Some are independent of $k$ (ignoring $\la_0$), while others have an index
$j>0$ for which the Dynkin label $\la_j$ grows linearly with $k$. The
quantum dimensions of the first kind of candidates converge as $k \rightarrow
\infty$ to the corresponding Weyl dimensions, while
the quantum dimensions of the second kind of candidates will all tend to
infinity. We will consider the two kinds of candidates separately; in
this Step 3 we first address those independent of $k$. The quantum dimensions
of the final four candidates in Lemma 2, all for $E_{8,15}$,
can be explicitly computed, and are all found to be far larger than
$\D_{15}(\om^1)$. All other $k$-independent candidates are of the form
$\om^i$. For the classical algebras, this step permits us to complete
the proof of the Proposition.

\smallskip
Let $\la,\mu$ be independent of $k$, and lie in $P_{+}(X_{\ell,k_0})$. Then
directly from (3.3) we find (a similar calculation was done in [10])
$$
{\partial\over \partial k}{\D_k(\la) \over \D_k(\mu)} = {\D_k(\la) \over
\D_k(\mu)} \, E_k(\la+\rho,\mu+\rho)\ ,
\eqno(4.6a)
$$
where
$$
E_k(\be,\ga):={\pi\over n^2}\sum_{\al>0}\Big[\ga\cdot\al\,\cot\Big(
\pi\,{\ga\cdot\al\over n}\Big)-\be\cdot\al\,\cot\Big(\pi\,{
\be\cdot\al\over n}\Big)\Big]\ .
\eqno(4.6b)
$$
{}From (4.6a), we find that if $\D_{k_0}(\la) \ge \D_{k_0}(\mu)$, and
$E_k(\la+\rho,\mu+\rho) > 0$ for all $k\ge k_0$, then $\D_k(\la) >\D_k(\mu)$
for all levels $k>k_0$.  Thus we begin by verifying the following, for all
$k\ge 1$:

\leftskip 1.5cm
\item{(i)} for $B_\ell$ and $\ell\ge 4$ \ :\ \
$E_k(\om^\l+\rho,\om^1+\rho)>0$,
\item{(ii)} for $C_{\ell}$ and $\ell\ge 2$ \ :\ \
$E_k(\om^\l+\rho,\om^1+\rho)>0$,
\item{(iii)} for $D_\ell$ and $\ell\ge 5$ \ :\ \
$E_k(\om^\l+\rho,\om^1+\rho)>0$,
\item{(iv)} for $E_6$ \ :\ \  $E_k(\om^i+\rho,\om^1+\rho)>0$ for $i=2,3,6$,
\item{(v)} for $E_7$ \ :\ \  $E_k(\om^i+\rho,\om^6+\rho)>0$ for all $i\neq 6$,
\item{(vi)} for $E_8$ \ :\ \  $E_k(\om^i+\rho,\om^1+\rho)>0$ for all $i\neq 1$,
\item{(vii)} for $F_4$ \ :\ \  $E_k(\om^i+\rho,\om^4+\rho)>0$ for all $i\neq
4$,
\item{(viii)} for $G_2$ \ :\ \  $E_k(\om^1+\rho,\om^2+\rho)>0$.

\leftskip 0cm
\smallskip \noindent
That the result (iii) does not hold for $\l=4$ is expected since
$\om^4\in[\om^1]$ there, and is of no consequence. On the other hand,
$B_3$ missing from (i) means it will have to be treated
separately.

We will illustrate how to obtain (i)--(viii), by working out $B_\l$
explicitly. Defining $c_i(x)=|\{\al > 0 \;|\; \al \cdot (\rho+\om^i)=x\}|$, we
find for $B_\l$:
$$
c_1(x) - c_\l(x) = \cases{
-2 & if $x=1$, \cr
-1 & if $3 \leq x \leq 2\l-3$ is an odd integer, \cr
1 & if $2x \neq 2\l-1$ is an odd integer between 1 and $2\l+1$, \cr
0 & otherwise. \cr}
\eqno(4.7)
$$
Setting $f(x):={\pi x\over n^2} \cot{\pi x\over n}$, we deduce from (4.7)
that for all $k\ge 1$ and $\l\ge 4$,
$$\eqalignno{
E_k(\om^\l+\rho,\om^1+\rho) =&-f(1) + f(\l+\half) +
\sum_{j=0}^{\l-2}\,\big\{f(j+\half)-f(2j+1)\big\} &\cr
=& \big\{f({5 \over 2})-f(3)\big\} - \big\{f(1)-f({3 \over 2})\big\} +
\big\{f(\half)-f(1)\big\} &\cr
& + \sum_{j=3}^{\l-2}\, \big\{f(j+\half)-f(2j-1)\big\} +
\big\{f(\l+\half)-f(2\l-3)\big\}. &(4.8)\cr}
$$
The difference of the first two braces is strictly positive
because the
function $f(x)$ is concave over $[0,n[$, while the other terms are positive
since $f(x)$ decreases over $[0,n[$. Thus $E_k(\om^\ell+\rho,\om^1+\rho)>0$
in this case. The other $X_\l$ are done similarly.

\smallskip
Next, we will find a $k_0$ such that all $\om^i\in P_+(X_{\l,k_0})$, and
$\D_{k_0}(\om^f)\le\D_{k_0}(\om^i)$. For the exceptional algebras this
is easy: we just explicitly compare the quantum dimensions $\D_k(\om^i)$
for the small levels $k\ge{\rm max}_j\{a_j^\vee\}$.
We find, for $E_6$, $E_7$, $E_8$, $F_4$, and $G_2$,
that $k_0=3$, 4, 6, 4, and 3, respectively.

For $X_\ell=B_\l$ ($\l>3$) and $D_\l$ ($\l>4$), it suffices to note that
at $k=1$, $\om^f\in[0]$ and $\om^\l\in P_+(X_{\l,1})$.
For $X_\l=C_\l$, it suffices to compute the level 2 quantum
dimensions, which is easy to do  from rank--level duality:
$$
{\D_2(\om^\ell)\over \D_2(\om^1)}=\left({1 \over
2\sin({\pi \over 2\l+6})\,\sin({3\pi \over 2\l+6})}\right)\,/\,\left(
4\cos({\pi \over 2\l+6})\,\cos({3\pi \over 2\l+6})\right)=
{1 \over 2\sin({\pi \over \l+3})\,\sin({3\pi \over \l+3})}.
\eqno(4.9)
$$
We find from (4.9) that $\D_2(\om^\l) \geq \D_2(\om^1)$
for all $\l \geq 3$, with equality only if $\l=3$. One more
calculation then shows that for $C_2$, $\D_3(\om^1) = \D_3(\om^2)$.

Finally for $B_3$, the quantity $E_k(\om^3+\rho,\om^1+\rho)$, as given on the
first
line of (4.8), is negative for all $k \geq 1$, so that
$\D_k(\om^3)/\D_k(\om^1)$
decreases with $k$. But since its value tends to 8/7 as $k \rightarrow
\infty$ (the ratio of the Weyl dimensions), it is bigger than 1 for all $k$.

Hence from (i)--(iii) above, together with the results of the
previous step, one obtains that $[\om^1]$ has the unique smallest quantum
dimension among the $[\om^i]$ for $B_{\l,k}$ and $D_{\l,k}$, $k \geq 3$, and
also for $C_{\l,k}$, $\l,k \geq 2$ and $\l + k \geq 6$.

\smallskip
This immediately concludes the proof of the Proposition for $D_{\l,k}$,
$k>2$, but in fact is also enough to complete the proof for $B_{\l,k}$ and
$C_{\l,k}$. For $C_{\l,k}$, the rank--level duality described in the previous
step implies
$$
\D(k\om^1) = \D'(\om'^k) > \D'(\om'^1) = \D(\om^1),
\eqno(4.10)
$$
for all $\l,k \geq 2$ and $\l + k \geq 6$. When $k=1$, $k\om^1 =\om^1$ and
$\om^\l \in [0]$. The remaining cases $C_{2,2}$, $C_{2,3}$ and $C_{3,2}$ can be
checked explicitly with the results given in Table 3.

For $B_\l$, and $k>6$ odd, rank--level duality implies
$$
\D(k\om^\ell)=\sqrt{2}\,\D'(\om'{}^{(k-1)/2})>\sqrt{2}\,\D'(\om'{}^1)=
\sqrt{2}\,\D(\om^1)>\D(\om^1).
\eqno(4.11)
$$
The same applies when $k>6$ is even. For $k=3$, we can explicitly
compute all quantum dimensions, using rank--level duality; we find the result
indicated in Table 3. For $3<k\le 6$, it suffices to show $\D(\om^1)<\D(k
\om^\ell)$ --- again, rank--level duality is the most efficient means. For
example,
$$
\D_4(\om^1)={\sin^2({2\pi\over 2\ell+3})\over \sin^2({\pi\over 2\ell+3})},
\qquad \D_4(4\om^\ell)=2{\sin({2\pi\over 2\ell+3})\over \sin({\pi\over 2\ell+3}
)}.
\eqno(4.12)
$$

\bigskip \noindent
{\it Step 4.} All that remains is to compare $\D_k(\om^f)$
with $\D_k(\la^k)$ for those candidates $\la^k$ of the exceptional algebras
which depend explicitly on $k$. For each $\la^k$, there exists a unique
Dynkin index $j>0$ such that $(\la^k)_j$ grows like $k/a^\vee_j$. For each
$\la^k$, we will consider $\D_k(\la^k)$ separately for each congruence
class of $k$ modulo $a_j^\vee$. Then $\D_k(\la^k)$ along such a congruence
class can be written  as a product of
$$
g_{\al\be\ga}(n):={\sin(\pi \,(\al+\be/n))\over \sin(\pi\,\ga/n)},\qquad
{}.
\eqno(4.13a)
$$
for $\al,\be,\ga$ independent of $n$ and obeying the inequalities
$$0\le \al \le {1\over 2},\quad 0<\al+\be/n<1,\quad 0<\ga <n.\eqno(4.13b)$$
Now, $g_{\al\be\ga}(n)$ is an increasing function of $n\ge 0$ if $\be<0$ or
$\al=1/2$, or of $n\ge 2\be$ if
$\be>\ga$. Also, for $0<\al<{1\over 2}$, $g_{\al
\be\ga}(n)$ is an increasing function of
$$
n\ge {\rm max}\left\{{\ga-\be\over \al},\
2\ga\right\}.
\eqno(4.14)
$$
These lower bounds for $n$ suffice to reduce the
proof of the Proposition for the exceptional algebras to a finite computer
search.

\smallskip
Write $dim(\om^f)$ for the Weyl dimension of the representation of $X_\ell$
with highest weight $\om^f$. The strategy is to use these simple results
concerning when $g_{\al\be\ga}(n)$ is increasing with $n$, to find a level
$k_0$ such that $\D_k(\la^k)$ is increasing (along each congruence class
of $k$) for $k\ge k_0$. Running through all $k$-dependent candidates and
their congruence classes, we obtain the following
ranges for $k_0$: 2 to 2 for $G_2$; 7 to 7 for $F_4$; 10 to 10 for $E_6$;
14 to 14 for $E_7$; and 28 to 29 for $E_8$. Explicitly computing $\D_k(\la^k)$
for $k<k_0$, we find that in fact for each $\la^k$, $\D_k(\la^k)$ is
monotonically increasing along each congruence class of $k$ modulo $a_j^\vee$.

Now for each $\la^k$ and each congruence class of $k$, let $k_1$ be the first
level satisfying $dim(\om^f)\le\D_{k_1}(\la^{k_1})$. For $k_1$, we get the
following
ranges: 5 to 6 for $G_2$; 5 to 7 for $F_4$; 5 to 7 for $E_6$; 5 to 13 for
$E_7$; and 7 to 31 for $E_8$.

We know from (4.6a) that
$\D_k(\om^f)$ is monotonically
increasing; by the Weyl dimension formula it converges to $dim(\om^f)$.
Therefore we know $\D_k(\om^f)<\D_k(\la^k)$ for all $k\ge k_1$. The remaining
finitely many $k$ can then be explicitly checked on a computer.

\vskip 1truecm
\noindent{\bf 5. The Classical Algebras}

\vskip 0.7truecm \noindent
In this section, we proceed to detail steps 2 and 3 of the proof of the
Theorem, as outlined
in section 2, for the four series of classical simple Lie algebras, with the
exceptions of $B_{\l,2}$ and $D_{\l,2}$ which we consider in section 6. In each
case, we first recall the relevant Lie algebraic data, and then explicitly
give all automorphism invariants. The fusion products we need are computed
using (3.6c).

\vskip 0.5truecm \noindent
{\it 5.1. The A--Series}

\medskip
All colabels $a_i^\vee$ are equal to 1, so that $h^\vee=\l+1$ and a weight
of $P_+(A_{\l,k})$ satisfies $\la_0+\la_1+\dots+\la_\l=k$. The charge
conjugation $C$ acts as $C(\la)=(\la_0;\la_\l,\la_{\l-1},\ldots,\la_1)$ and is
trivial if $\l=1$.

The simple currents form a cyclic group of order $\l+1$, generated by $J$ with
action $J(\la) = (\la_\l;\la_0,\la_1,\ldots,$ $\la_{\l-1})$, corresponding to a
rotation of the extended Dynkin diagram. Their charge and conformal weight are
equal to $Q_{J^m}(\la) = {m \over \l+1} \sum_{j=1}^\l \, j\la_j$ and
$h_{J^m(0)} = km(\l+1-m)/2(\l+1)$.

\smallskip
Choose any positive integer $m$ dividing $\l+1$, such that $k(\l+1)/m$ and $m$
are coprime if $m$ is odd, and such that $k(\l+1)/2m$ is an integer coprime
with $m$ if $m$ is even. In both cases, this means that we can find an integer
$v$ such that $vk(\l+1)/2m \equiv  1 \bmod m$. To each such divisor $m$ of
$\l+1$, one associates the automorphism invariant given by
$$
\si_m(\la) = J^{-v(\l+1)^2Q_J(\la)/m}(\la),
\eqno(5A.1)
$$
and first found in [13].
These and their conjugations $C\circ\si_m$ are the automorphisms appearing in
Table 1; that they form the complete set is proved in [17].
The total number of different automorphism invariants equals  $2^{c+p+t}$,
where
$$\eqalign{
& c = \cases{ -1 & if $\l=1$ and $k=2$, \cr
               0 & if $\l=1$ and $k \neq 2$, or $\l \geq 2$ and $k \leq 2$, \cr
               1 & otherwise; \cr} \cr
& p = \hbox{number of distinct odd primes which divide $\l+1$ but not $k$;} \cr
& t = \cases{ 0 & if either $\l$ is even, or $\l$ is odd and $k\equiv 0 \bmod
                            4$, \cr
                & or $\l\equiv 1 \bmod 4$ and $k$ is odd, \cr
              1 & otherwise. \cr} \cr}
\eqno(5A.2)
$$
All $A_{\l,k}$ automorphism invariants have order 2 and commute.

\vskip 1.5truecm \noindent
{\it 5.2. The B--Series}

\medskip
A weight in $P_+(B_{\l,k})$ satisfies $\la_0+\la_1+2\la_2+\ldots+2\la_{\l-1}
+\la_\l = k$, and the dual Coxeter number of $B_\l$ is $h^\vee=2\l-1$. As $B_2
\cong C_2$, we take $\l \geq 3$.

The charge conjugation $C$ is trivial, but there is a simple current of order
2, which exchanges the zero--th and first components, $J(\la)=(\la_1;\la_0,
\la_2,\ldots,\la_\l)$. It has $Q_J(\la)=\la_\l/2$ and $h_{J(0)}=k/2$. When $k$
is odd, there is the simple current automorphism invariant [3]
$$
\si_J(\la) = J^{\la_\l}(\la), \quad \hbox{ for $k$ odd.}
\eqno(5B.1)
$$
As reported in Table 1, this is the only non--trivial invariant for $k \neq 2$,
whereas for $k=2$, there are a number of exceptional invariants.
As already apparent in Table 3, $k=2$ is very special, and we
defer its full description to the next section. The case $k=1$ is
straightforward (see [15]). We proceed here with the proof
when $\l \geq 3$ and $k \ge 3$.

\medskip
{}From the corollary of section 4, we know that the action of any automorphism
on the first fundamental weight is necessarily of the form
$\si(\om^1)=J^b(\om^1)$. Suppose $b=1$. Then the norm condition yields
$$
(\rho + J\om^1)^2 - (\rho + \om^1)^2 \equiv  (k-2)n \equiv  0 \; \bmod 2n.
\eqno(5B.2)
$$
Therefore $\si(\om^1)=J(\om^1)$ requires $k$ to be even.

\medskip
The basic idea of the proof is the same as for the $A_\l$ series in [17],
but with the
extra complication that not all fundamental representations are contained in
fusion powers of $\om^1$. Thus we need a second weight,
for which a convenient choice is the spinor $\om^\l$. The full proof (for $k
\neq 2$) includes three steps:

\item{\it (i)} we first show that an automorphism which fixes $\om^1$ and
$\om^\l$ is necessarily trivial (this result also holds for $k=2$);
\item{\it (ii)} assuming that $\om^1$ is fixed, we find only four
possibilities for $\si(\om^\l)$ consistent with the action of $\si$ on the
fusion product
$\psi \times \om^\l$ (here $\psi=\om^2$ is the adjoint representation); from
this, we easily conclude that the only globally acceptable solutions are
$\si_1$ (all $k$) and $\si=\si_J$ ($k$ odd); \item{\it (iii)} finally, we show
that the assumptions $k$ even and $\si(\om^1)=J(\om^1)$ are not compatible with
$\si$ being an automorphism of the fusion ring.

\smallskip
We first of all introduce the orthogonal basis $\{e_i\}$, convenient for
computing
fusion products. So we will set $\la=[x_1,x_2,\ldots,x_\l]$, with the
orthogonal
components given in terms of the Dynkin components by
$$\eqalignno{
& x_i = \la_i + \ldots + \la_{\l-1} + {\la_\l \over 2}, &(5B.3a) \cr
& x_\l ={\la_\l \over 2}. &(5B.3b) \cr}
$$
In this basis, the metric is the identity $\la \cdot \la' = \sum_i x_i
x'_i$, and the Weyl vector is $\rho=(1,\ldots,1)= [\l-\half,\l-{3 \over 2},
\ldots,{3 \over 2},\half]$.

\vskip 0.5truecm \noindent
{\it (i)}\\
We start off by proving that if $\om^1$ and $\om^\l$ are both fixed
by $\sigma$, then all weights are fixed, so that $\si = \si_1$. The
weights of the defining representation $\om^1$ are $\{0, \; \pm
e_i\}_{1 \leq i \leq \l}$, so that
$$\eqalignno{
& \om^1 \times \om^1 = 0 + \om^2 + (2\om^1), &(5B.4a) \cr
& \om^1 \times \om^i = \om^{i-1} + \om^{i+1} + (\om^1 + \om^i), \quad
\hbox{ for $2 \leq i \leq \l-2$.} &(5B.4b) \cr}
$$
The norms of the weights appearing in (5B.4) read $(\rho + \om^i)^2 = \rho^2 +
i(2\l+1-i)$ for $1 \leq i \leq \l-1$, $(\rho + 2\om^1)^2 = \rho^2 + 4\l+2$ and
$(\rho + \om^1 + \om^i)^2 = \rho^2 + i(2\l+1-i) + 2\l+2$, also for $1 \leq i
\leq \l-1$. Assuming $\si(\om^1)=\om^1$, we obtain that
$\si$ must permute the weights on the r.h.s. of (5B.4a). But a
non--trivial permutation is forbidden by the values of their norms, so
that $\si(\om^2)=\om^2$. The same argument applies to (5B.4b) with $i=2$,
showing that
$\om^3$ must be fixed by $\si$, and by induction, all weights $\om^i$, $i<\l$,
must be fixed. If $\om^\l$ is assumed to be fixed as well, then Lemma 1
implies that the whole of the alc\^ove is fixed, and that $\si=\si_1$.

\vskip 0.5truecm \noindent
{\it (ii)}\\
Here we assume that $\om^1$ is fixed by $\si$, and show that the only
automorphisms with this property are $\si=\si_1$ and, for $k$ odd, $\si=\si_J$.

\smallskip
The fusion (5B.4a) shows that the adjoint $\psi = \om^2 = [1,1,0,\ldots,0]$
must be fixed by $\si$. We
first compute the fusion of $\psi$ with the spinor $\om^\l =
[\half,\ldots,\half]$,  then compare it with that of $\psi$ with
$\sigma(\om^\l)$ and require they be compatible.

\smallskip
The weights of the spinor representation $\om^\l$ are $P(\om^\l)=\{[\pm
\half,\ldots,\pm
\half]\}$ (with uncorrelated signs), so that the weights appearing in
$\psi \times \om^\l$ have the following form: {\it (a)} in the first two
positions, there will be $\half$'s and $3 \over 2$'s, but a $\half$
followed by a $3 \over 2$ puts the weight in a wall of the alc\^ove (meaning
it would be fixed by a Weyl reflection and so does not contribute), and
{\it (b)} in the
last $\l-2$ positions, there will be $\half$'s and $-\half$'s, but a $-\half$
followed by a $\half$ or an ending $-\half$ also puts the weight in a
wall (recall that we are using weights non--shifted by $\rho$). Thus
$$
\psi \times \om^\l = \om^\l  + [{3 \over 2},\half,\ldots,\half]
+ [{3 \over 2},{3 \over 2},\half,\ldots,\half].
\eqno(5B.5)
$$
Set $\la=\sigma(\om^\l)$. The weight diagram of the adjoint is the set of
roots of $B_\l$ so that
$$
P(\psi) + \la =\{ \la \pm e_i,\; \la \pm (e_i-e_j),\; \la \pm (e_i+e_j),\;
\la\}_{1\le i<j\le\l}.
\eqno(5B.6)
$$
{}From (5B.5) and (3.8d), we require that $N_{\psi,\la}^\la=1$. But
mult${}_\psi(0)=\l$, and this implies from (3.6c) that there should be
$\l-1$ non--zero roots $\alpha$ such that the weights $\la - \alpha$
get out of the alc\^ove and brought back onto $\la$ by an odd Weyl
transformation. Looking at all non--zero roots, we find that those which can
take $\la - \alpha$ out of the alc\^ove and off the walls are

\item{1.} the $\l+1$ affine simple roots $\alpha_i$ ($\alpha_0=-\psi$)
iff $\la \cdot \alpha_i = 0$ for $i \geq 1$ ({\it i.e.} the $i$--th Dynkin
label
equal to zero), and $\la \cdot \psi = k$ for
$i=0$. One easily checks that $w_i(\rho+\la-\alpha_i)=\rho+\la$ with $w_i$ the
Weyl reflector through the $i$--th hyperplane. [For $i=0$, the reflection is
given by $w_0(\la) = \la + (n - \la \cdot \psi)\psi$.]
\item{2.} the roots $\alpha = \pm e_i + e_\l$ for $1 \leq i \leq \l-1$ iff $\la
\cdot \alpha_\l = x_\l = 0$. In this case, we have $w_\l(\rho+\la-\alpha)
= \rho+\la \mp e_i$, so that the weights $\la \pm e_i - e_\l$ and
$\la \pm e_i$ all cancel out.

\smallskip
Since the condition $N_{\psi,\la}^\la=1$ requires that $(\l-1)$ $\la$'s cancel
against some $\la-\alpha_i$ for some choice of $(\l-1)$ affine simple roots
$\alpha_i$, we find that $\la$ must have either $\l-1$ zero Dynkin labels and
satisfy $\la \cdot \psi = x_1 + x_2 < k$, or else $\l-2$ zero Dynkin labels and
satisfy
$\la \cdot \psi = x_1 + x_2 = k$. In addition, the fusion $\psi \times \la$
must
contain exactly three weights. From (5B.6), the result is that these
two conditions, $N_{\psi,\la}^\la = 1$ and
$\sum_\mu \; N_{\psi,\la}^\mu = 3$, force $\la$ to be one of the
following four weights
$$
\la = \si(\om^\l) \in \big\{ \om^1,\; J(\om^1), \; \om^\l, \; J(\om^\l)\big\}.
\eqno(5B.7)
$$

The first weight in (5B.7) must be discarded since it was assumed to be fixed
by $\si$. If $\la=J(\om^1)$, then using $N_{\lambda,J(\mu)}^{J(\nu)} =
N_{\lambda,\mu}^\nu$, a straightforward consequence of $S_{\la,J(\la')} = {\rm
e}^{2i\pi  x_\l} \,S_{\la,\la'}$ and the Verlinde formula, we obtain from
(5B.4a)
$$
\om^1 \times J(\om^1) = [k-2,0,\ldots,0] +[k-1,1,0,\ldots,0]
+ [k,0,\ldots,0],
\eqno(5B.8)
$$
which must be the transform by $\si$ of
$$
\om^1 \times \om^\l = \om^\l + [{3 \over 2},\half,\ldots,\half],
\eqno(5B.9)
$$
clearly impossible.

The solution $\la=\si(\om^\l)=\om^\l$ leads to $\si=\si_1$ by step {\it (i)},
since
$\si$ then fixes both $\om^1$ and $\om^\l$. The remaining possibilitiy
$\la=J(\om^\l)$ requires $k$ odd for norm reasons, and leads to the simple
current automorphism $\si=\si_J$ of (5B.1). Indeed $\si_J^{-1} \circ \si$,
fixing $\om^1$ and $\om^\l$, must be trivial, implying $\si = \si_J$.

\vskip 0.5truecm \noindent
{\it (iii)}\\
We now assume that $k$ is even and that $\si(\om^1) =
J(\om^1)$. With the
results of step {\it (ii)}, it is easy to show that these assumptions lead to a
contradiction: for $k \geq 4$ even, we will find that there is no automorphism
such that $\si(\om^1)=J(\om^1)$.

\medskip
The identity $N_{J(\lambda),J(\mu)}^\nu = N_{\lambda,\mu}^\nu$ implies from
(5B.4a)
$$
\om^1 \times \om^1 = J(\om^1) \times J(\om^1) = 0 + \om^2 + (2\om^1).
\eqno(5B.10)
$$
As before, we obtain that the adjoint
$\psi$ must be fixed, since $\si$ must preserve the r.h.s. of (5B.10). The
argument we used in the second part {\it (ii)} then shows that
$\la=\si(\om^\l)$ must be one of the four weights in (5B.7).

\smallskip
Take first $\la=\om^1$. This again implies that the fusion product (5B.8) must
be the $\si$--transform of that in (5B.9), which is impossible. The
second weight $\la=J(\om^1)$ in (5B.7) must also be discarded since
$J(\om^1) = \si(\om^1)$ is already the image of $\om^1$.

If $\la=\om^\l$, we obtain from (5B.9), using once more $N_{J(\la),\mu}^
{J(\nu)} = N_{\lambda,\mu}^\nu$,
$$
J(\om^1) \times \om^\l = [k-\half,\half,\ldots,\half] + [k-{3 \over
2},\half,\ldots,\half].
\eqno(5B.11)
$$
Again (5B.11) must be the image under $\si$ of (5B.9). This requires that
$\om^\l
= [\half,\ldots,\half]$ be in the r.h.s. of (5B.11), implying $k=1$ or $k=2$,
contrary to the assumption $k \geq 4$.

Finally, $\la=J(\om^\l)$ requires $k$ odd.

Therefore, all four possibilities in (5B.7) lead to a contradiction, and the
proof of the Theorem is complete for the $B_{\l,k}$ algebras, $k \neq 2$.
\cqfd

\vskip 0.5truecm \noindent
{\it 5.3. The C--Series}

\medskip
A weight of $P_+(C_{\l,k})$ satisfies $\la_0+\la_1+\ldots+\la_\l=k$ and the
dual Coxeter number is equal to $h^\vee=\l+1$. Here too, the charge conjugation
$C$ is trivial, and there is one simple current $J$, of order 2,
defined by $J(\la)=(\la_\l;\la_{\l-1},\ldots,\la_1,\la_0)$. It has
$h_{J(0)}=k\l/4$ and $Q_J(\la)=\sum_{j=1}^\l \, j\la_j/2$.

When $k\l\equiv 2 \bmod 4$, there is a simple current automorphism invariant
given by [3]
$$
\si_J(\la) = J^{2Q_J(\la)}(\la), \qquad \hbox{if $k\l\equiv 2 \bmod 4$.}
\eqno(5C.1)
$$
The diagonal invariant $\si_1$ and $\si_J$ are the only automorphism invariants
(note that for $\l=2$ and $k=1$, $\si_J =\si_1$).

\medskip
Let $\si$ be any automorphism invariant of $C_{\l,k}$. From the corollary of
section 4, we have that, for any $k$, $\si(\om^1)=J^b(\om^1)$ for some
$b=0,1$. Suppose $b=1$. Then the norm condition yields
$$
(\rho + \om^1)^2 \equiv (\rho + J\om^1)^2 \bmod 2n \quad \Rightarrow \quad
\half(k\l-2)n \equiv  0 \bmod 2n.
\eqno(5C.2)
$$
Therefore $\si(\om^1)=J(\om^1)$ requires $k\l\equiv 2 \bmod 4$. But precisely
for those values of $k$ and $\l$, there exists the automorphism invariant
$\si_J$, whose action on $\om^1$ is also $\si(\om^1)=J(\om^1)$. Thus replacing
$\si$ by
$\si_J^{-1} \circ \si$, we may assume for all $k$ that $\si(\om^1)=\om^1$, and
show that the only such automorphism is trivial. This will complete the proof
for the $C_\l$ series.

\smallskip
 The fusion of $\om^1$ with the other fundamentals reads
$$\eqalignno{
& \om^1 \times \om^1 = 0 + \om^2 + (2\om^1), &(5C.3a) \cr
& \om^1 \times \om^i = \om^{i-1} + \om^{i+1} + (\om^1 + \om^i), \quad
\hbox{for $2 \leq i \leq \l-1$}. &(5C.3b) \cr}
$$
The norms of the weights in the r.h.s. of (5C.3a--b) are equal to $(\rho +
\om^i)^2 = \rho^2 + i(\l+1-{i \over 2})$, $(\rho+2\om^1)^2 = \rho^2 + 2\l+2$,
and $(\rho+\om^1+\om^i)^2 = \rho^2 + i(\l+1-{i \over 2}) + \l + {3 \over 2}$.
Assuming that $\om^1$ is fixed by $\si$, the norm argument shows from
(5C.3a) that $\om^2$ must be fixed, and then from (5C.3b), that all $\om^i$ are
fixed as well. In turn, Lemma 1 implies that all weights of
$P_+(C_{\l,k})$ must be fixed, and $\si=\si_1$. \cqfd

\vskip 0.5truecm \noindent
{\it 5.4. The D--Series}

\medskip
A weight of $P_+(D_{\l,k})$ satisfies $\la_0+\la_1+2\la_2+\ldots+2\la_{\l-2}+
\la_{\l-1}+\la_\l=k$, and the height is $n=k+2\l-2$. Since $D_3 \cong A_3$, we
will assume $\l \geq 4$.

For any $\l$, there is the outer automorphism
$$
C_1(\la) = (\la_0;\la_1,\ldots, \la_{\l-2},\la_\l,\la_{\l-1}).
\eqno(5D.1a)
$$
For $\l$ odd, $C=C_1$ is the charge conjugation, while for $\l$ even, the
charge conjugation is trivial. Moreover when $\l=4$, there are four new outer
automorphisms given by
$$
C_2(\la)=(\la_0;\la_4,\la_2,\la_3,\la_1), \quad C_3=C_1C_2, \quad
C_4=C_2C_1, \quad C_5=C_1C_2C_1.
\eqno(5D.1b)
$$
Together with $C_0=\si_1$, these six $C_i$ correspond to the different
permutations of the Dynkin labels $\la_1,\la_3,\la_4$.

\smallskip
There are three non--trivial simple currents, $J_v$, $J_s$ and $J_c=J_v \circ
J_s$. Explicitly, we have
$$\eqalignno{
& J_v\la=(\la_1;\la_0,\la_2,\ldots,\la_{\ell-2},\la_\ell,\la_{\ell-1}),
& (5D.2a) \cr
& Q_v(\la)=(\la_{\ell-1}+\la_\ell)/2, \quad h_{J_v(0)}=k/2. &(5D.2b) \cr}
$$
The expressions for $J_s$ and $J_c$ depend on the parity of $\l$ and are given
by
$$\eqalignno{
& J_s(\la)=\cases{
(\la_\ell;\la_{\ell-1},\la_{\ell-2},\ldots,\la_1,\la_0) & if $\l$ is even, \cr
(\la_{\ell-1};\la_\ell,\la_{\ell-2},\ldots,\la_1,\la_0) & if $\l$ is odd,
\cr} &(5D.2c) \cr
& Q_s(\la)=\sum_{j=1}^{\ell-2}j\la_j/2-{\l-2 \over 4}\la_{\ell-1}
-{\l \over 4}\la_\ell, \quad h_{J_s(0)}=k\l/8, &(5D.2d) \cr}
$$
and
$$\eqalignno{
& J_c(\la)=\cases{
(\la_{\l-1};\la_\l,\la_{\ell-2},\ldots,\la_2,\la_0,\la_1) & if $\l$ is even,
\cr
(\la_\l;\la_{\l-1},\la_{\ell-2},\ldots,\la_2,\la_0,\la_1) & if $\l$ is odd,
\cr} &(5D.2e) \cr
& Q_c(\la)=\sum_{j=1}^{\ell-2}j\la_j/2-{\l \over 4}\la_{\ell-1}
-{\l-2 \over 4}\la_\ell, \quad h_{J_c(0)}=k\l/8. &(5D.2f) \cr}
$$
All three simple currents have order 2, except $J_s$ and $J_c$ which have
order 4 if $\l$ is odd. We denote by $N_s$ the order of $J_s$ (equal to the
order of $J_c$).

\medskip
Corresponding to these simple currents, one defines the following simple
current automorphism invariants
$$\eqalignno{
& \si_v(\la) = J_v^{\la_{\ell-1}+\la_\ell}(\la), \qquad\quad\ \hbox{if $k$
is odd,} &(5D.3a) \cr
& \si_s(\la) = J_s^{N_s^2k\ell\,Q_s(\la)\,/8}(\la), \qquad \hbox{if
$N_sk\l \equiv  8 \bmod 16$,} &(5D.3b) \cr
& \si_c(\la) = J_c^{N_s^2k\ell\,Q_c(\la)\,/8}(\la), \qquad
\hbox{if $N_sk\l \equiv 8 \bmod 16$,} &(5D.3c) \cr}
$$
with $\si_s=\si_c$ if $\l$ is odd. The automorphism invariant $\si_v$ was
found in [3] as well as $\si_s$ and $\si_c$ for $\l$ even, while
$\si_s$ for $\l$ odd was discovered in [1].

The last simple current automorphism invariant for $D_{\ell,k}$, when $k$ and
$\ell$ are both even and $k\ell \equiv 0 \bmod 8$, was found in [30]. It is
the integer spin simple current automorphism we mentioned in section 3, and
explicitly reads
$$\si_{vsc}(\la)=\cases{
\la & if $Q_v(\la) \equiv 0$, $Q_s(\la) \equiv 0 \bmod 1$, \cr
\noalign{\smallskip}
J_v(\la) & if $Q_v(\la) \equiv 0$, $Q_s(\la) \equiv \half \bmod 1$, \cr
\noalign{\smallskip}
J_s(\la) & if $Q_v(\la) \equiv \half$, $Q_s(\la) \equiv 0 \bmod 1$, \cr
\noalign{\smallskip}
J_c(\la) & if $Q_v(\la) \equiv \half$, $Q_s(\la) \equiv \half \bmod 1$. \cr}
\eqno(5D.3d)
$$

Obviously any product of these with each other (when the values of $k$
and $k\ell$ allow it) and with the ${C}_i$ will define other automorphism
invariants. Together, they generate all of them, for $k\ne 2$.

When $k\equiv \l\equiv 2 \bmod 4$, $\si_s$ and $\si_c$ generate an
Abelian subgroup of order 4, containing the elements $\si_1,\si_s,\si_c$, and
$\si_s \circ \si_c=\si_c \circ \si_s$. In this case the automorphism
invariants are just $C_1^a\si_s^b\si_c^c$, where $a,b,c=0,1$.

When $\l\equiv 4 \bmod 8$ and $k$ is odd, the subgroup generated by $\si_s$
and $\si_c$ is of order 6, and consists of
the elements $\si_1,\si_s,\si_c,\si_s \circ \si_c,\si_c \circ \si_s$, and
$\si_s
\circ \si_c \circ \si_s = \si_c \circ \si_s \circ \si_c=\si_v$. Any
automorphism invariant in this case will look like $C_j \si$, where
$\si \in \langle \si_s,\si_c \rangle$ and $C_j$ is one of the 2 ($\l \neq 4$)
or
6 ($\l=4$) conjugations.

In general we have $C_1\si_v=\si_vC_1$, $C_j\si_{vsc}=\si_{vsc}C_j$
and $C_1\si_s=\si_cC_1$. Also, $\si_v^2=\si_s^2=\si_c^2=\si_{vsc}^2=\si_1$.

When $k=2$, there are in addition a number of exceptional invariants, detailed
in the next section, and first found in [12]. The proof for $k=1$ was done
in [15].

\bigskip
We now proceed to show, for $k \ge 3$, that this list of automorphism
invariants, also summarized in Table 1, is exhaustive. As usual, we first use
the results of section 4 to restrict the possible values of $\si(\om^1)$.

\medskip
Let $\si$ be any automorphism invariant of $D_{\l,k}$. From the corollary of
section 4, we have that, for any $k \neq 2$, $\si(\om^1) = C_jJ_s^aJ_v^b(
\om^1)$, where $C_j$ is some conjugation, and $a,b=0,1$. By replacing $\si$
with $C_j\circ \si$, we may drop $C_j$.

Consider first the possibility $a=1,b=0$. Then
$$
(\rho + \om^1)^2 \equiv (\rho + J_s\om^1)^2 \; \bmod 2n \quad \Rightarrow
\quad ({k\ell \over 4}-1)\,n \equiv 0 \; \bmod 2n.
\eqno(5D.4)
$$
Therefore $\si(\om^1)=J_s(\om^1)$ requires $k\ell \equiv 4 \bmod 8$. When
$\ell$
is even, there exists an automorphism invariant $\si_s$ for these $k,\ell$
with the property that $\si_s(\om^1)=J_s(\om^1)$; in this case, replacing
$\si$ with $\si_s \circ \si$, we may assume
$\si(\om^1)=\om^1$. On the other hand, when $\ell$ is odd, the charge
conjugation $C_1=S^2$ must commute with any automorphism invariant: {\it i.e.}
$C_1\circ \si \circ C_1=\si$. But this would be violated if $\si(\om^1)=J_s(
\om^1)$, since
$J_s(\om^1)=J_c(\om^1)$ only happens when $k=2$. Thus the case $a=1,b=0$ cannot
apply when $\ell$ is odd.
The identical argument applies to $a=b=1$.

Finally, consider $a=0,b=1$. In this case
$$
(\rho + \om^1)^2 \equiv (\rho + J_v\om^1)^2 \; \bmod 2n \quad \Rightarrow
\quad (k-2)n \equiv 0 \; \bmod 2n.
\eqno(5D.5)
$$
Therefore $\si(\om^1)=J_v(\om^1)$ requires $k$ even. If in addition $k\ell
\equiv 2 \bmod 4$, then replace $\si$ with $\si_s\circ\si$. If instead $\ell$
is also even, and $k\ell \equiv 0 \bmod 8$, then replace $\si$ with $\si_{vsc}
\circ\si$. Finally, if $\ell$ is even and $k\ell \equiv  4 \bmod 8$, then
replace $\si$ with $\si_s\circ\si_c\circ\si$. So in all cases, after
composing it with adequate automorphisms, one may assume $\si(\om^1)=\om^1$
{\it except} if both $k\equiv 0 \bmod 4$ and $\l$ is odd, in which case
$\si(\om^1)=J_v(\om^1)$ remains a possibility.

\medskip
The idea of the proof is exactly the same as in the $B_\l$ series, and is only
slightly more complicated due to the larger number of outer automorphisms. More
precisely, for $k \neq 2$, we will go through the proof of the following three
points:

\item{\it (i)} an automorphism which fixes $\om^1$ and the spinor $\om^\l$
must be trivial (true even for $k=2$);
\item{\it (ii)} assuming that $\om^1$ alone is fixed, there are now twelve
possibilities for $\si(\om^\l)$ consistent with the action of $\si$ on the
fusion product
$\psi \times \om^\l$; apart from the trivial solution $\si=\si_1$, this will
imply that the only globally acceptable solutions are $C_1$ (all $k$) and
$\si_v, \, C_1\si_v$ (for $k$ odd);
\item{\it (iii)} finally, the assumptions $k\equiv 0 \bmod 4$, $\l$ odd and
$\si(\om^1)=J_v(\om^1)$ are not compatible with $\si$ being an automorphism of
the fusion ring.

\smallskip
Again we first introduce an orthogonal basis $\{e_i\}$ in the weight space, and
write $\la=[x_1,x_2,\ldots,x_\l]$ with the new components given in terms of the
Dynkin labels as
$$\eqalignno{
& x_i = \la_i + \ldots + \la_{\l-2} + \half (\la_{\l-1} + \la_\l), &(5D.6a) \cr
& x_{\l-1} = \half (\la_{\l-1} + \la_\l), &(5D.6b) \cr
& x_\l = \half (-\la_{\l-1} + \la_\l). &(5D.6c) \cr}
$$
In that basis, the metric is the identity $\la \cdot \la' = \sum_i x_i
x'_i$, and the Weyl vector reads $\rho=(1,\ldots,1)=[\l-1,\l-2,\ldots,1,0]$.

\vskip 0.5truecm \noindent
{\it (i)}\\
Let us show that if $\om^1$ and $\om^\l$ are both assumed to be fixed
by a $\sigma$, then all weights are fixed as well and $\si = \si_1$. The weight
diagram of the defining representation is the set $P(\om^1)=\{\pm e_i\}_{1
\leq i \leq \l}$, and we obtain the following fusions
$$\eqalignno{
& \om^1 \times \om^1 = 0 + \om^2 + (2\om^1), &(5D.7a) \cr
& \om^1 \times \om^i = \om^{i-1} + \om^{i+1} + (\om^1 + \om^i), \quad
\hbox{for $2 \leq i \leq \l-3$}, &(5D.7b) \cr
& \om^1 \times \om^\l = \om^{\l-1} + (\om^1 + \om^\l). &(5D.7c) \cr}
$$
The usual norm argument applies once more. The first fusion (5D.7a) implies
that $\om^2$ is fixed by $\si$ if $\om^1$ is fixed. Then (5D.7b) shows that all
fundamental weights $\om^i$, for $3 \leq i \leq \l-2$, are fixed.
Finally, assuming $\om^\l$ fixed, the last fusion forces $\om^{\l-1}$ to be
fixed as well. From Lemma 1, the conclusion follows that all weights in
$P_+(D_{\l,k})$ are invariant, or $\si=\si_1$.

\vskip 0.5truecm \noindent
{\it (ii)}\\
We assume here that $\si(\om^1)=\om^1$ and classify all automorphisms with that
property, for $k \ge 3$.

\smallskip
The fusion (5D.7a) shows that the adjoint $\psi=\om^2=[1,1,0,\ldots,0]$ must
be fixed by any $\si$ which leaves $\om^1$ invariant. We will compute the
fusions $\psi \times \om^\l$ and $\psi \times \si(\om^\l)$ and require their
compatibility, thereby restricting $\si(\om^\l)$.

\medskip
The weight diagram of the spinor $\om^\l$ is $P(\om^\l) = \{[\pm
\half,\ldots,\pm
\half]\}$ where the number of $-$ signs is even. Arguing as in the $B_\l$ case,
we find
$$
\psi \times \om^\l = \om^\l  + [{3 \over 2},\half,\ldots,\half,-\half]
+ [{3 \over 2},{3 \over 2},\half,\ldots,\half].
\eqno(5D.8)
$$
Denote $\la=\sigma(\om^\l)$. The weight diagram of the adjoint is the set of
roots of $D_\l$ so that
$$
P(\psi) + \la = \{\la \pm (e_i-e_j),\; \la \pm (e_i+e_j),\; \la\}_{1\le i<j
\le \l}.
\eqno(5D.9)
$$
Again, mult${}_\psi(0)=\l$. As in $B_\l$, this implies that there must
be $\l-1$ non--zero roots
$\alpha$ such that $\la-\alpha$ gets out of the alc\^ove and is mapped back on
$\la$ by an odd Weyl transformation. In this case, we find that the only
non--zero roots which can take $\la-\alpha$ out of the alc\^ove
are the $\l+1$ affine simple roots $\alpha_i$, with $\alpha_0=-\psi$.
Moreover $\la-\alpha_i$ is out of the alc\^ove if and only if $\la \cdot
\alpha_i = 0$ for
$i \geq 1$, and $\la \cdot \psi = k$ for $i=0$. One also checks that
$w_i(\rho+\la-\alpha_i)=\rho+\la$ with $w_i$ the Weyl reflector through the
$i$--th hyperplane. Therefore, $N_{\psi,\la}^\la=1$ if and only if either
$\l-1$
Dynkin labels are zero and $\la \cdot \psi = x_1+x_2 < k$, or else $\l-2$
Dynkin
labels are zero and $\la \cdot \psi = x_1+x_2 = k$. The other condition we
obtain by comparing (5D.8) and (5D.9) is that the fusion $\psi \times \la$ must
contain exactly three weights. Altogether the two conditions $N_{\psi,\la}^\la
= 1$ and $\sum_\mu \, N_{\psi,\la}^\mu = 3$ force $\la=\si(\om^\l)$ to be one
of
the following twelve weights (given in the Dynkin basis)
$$\eqalignno{
& \la = \om^\l, \; \om^{\l-1}, \; (k-1,0,\ldots,0,1,0), \; (k-1,0,\ldots,0,1),
& (5D.10a) \cr
& \la = \om^1, \; (k-1,0,\ldots,0), & (5D.10b) \cr
& \la = (0,\ldots,0,1,k-1), \;(0,\ldots,0,k-1), \; (1,0,\ldots,0,k-1,0), \cr
& \quad \qquad \hbox{and } (0,\ldots,0,k-1,1),\; (0,\ldots,0,k-1,0), \;
(1,0,\ldots,0,k-1). &(5D.10c) \cr}
$$
It remains to examine these 12 possibilities case by case.

\medskip
The four weights in (5D.10a) correspond to $\la=\si(\om^\l)$ with $\si$ given
respectively by $\si=1$, $C_1$, $\si_v$ and $C_1
\si_v$ (the last two requiring $k$ odd for norm reasons). These four
automorphisms all leave $\om^1$ fixed, so that composing them with $\si$
leaves us with an automorphism which fixes both $\om^1$ and $\om^\l$, hence
trivial by step {\it (i)}. This shows that $\si=\si_1$, $C_1$, $\si_v$ and
$C_1\si_v$ everywhere.

\medskip
The possibility $\la = \om^1$ must be discarded since $\om^1$ was assumed to
be fixed. The second one, $\la=(k-1,0,\ldots,0)=J_v(\om^1)$, must also be
excluded. Indeed the identity $N_{\la,J_v(\mu)}^{J_v(\nu)}=N_{\la,\mu}^\nu$
applied to $\om^1 \times \om^1$ yields
$$
\om^1 \times J_v(\om^1) = J_v(0) + J_v(\om^2) + J_v(2\om^1).
\eqno(5D.11)
$$
This must be the image under $\si$ of the product $\om^1 \times \om^\l$ given
in (5D.7c), and which contains only two fields in its r.h.s., leading to a
contradiction.

\medskip
As to the six weights in (5D.10c), it is enough to consider the first three,
$\la^1:=(0,\ldots,0,1,$ $k-1)$, $\la^2:=(0,\ldots,0,k-1)$ and $\la^3: =
(1,0,\ldots,0,k-1,0)$, since the last three are their conjugates by $C_1$.
Let us compare the fusions of $\om^1$ with $\om^\l$ and with $\la^i\,$:
$$\eqalignno{
& \om^1 \times \om^\l = (\om^\l + e_1) + (\om^\l - e_\l), &(5D.12a) \cr
& \om^1 \times \la^1 = (\la^1 + e_\l) + (\la^1 - e_{\l-1}) + (\la^1 - e_\l),
&(5D.12b) \cr
& \om^1 \times \la^2 = (\la^2 + e_1) + (\la^2 - e_\l), &(5D.12c) \cr
& \om^1 \times \la^3 = (\la^3 - e_1) + (\la^3 + e_\l). &(5D.12d) \cr}
$$
Assuming $(\rho+\om^\l)^2 \equiv (\rho+\la^i)^2 \bmod 2n$ for consistency, one
obtains the norms
$$\eqalignno{
& (\rho + \om^\l - e_\l)^2 = (\rho + \om^\l)^2, &(5D.13a) \cr
& (\rho + \la^2 + e_1)^2 \equiv (\rho + \la^3 - e_1)^2 \equiv (\rho +
\om^\l)^2 + n \bmod 2n,  &(5D.13b) \cr
& (\rho + \la^2 - e_\l)^2 \equiv (\rho + \la^3 + e_\l)^2 \equiv (\rho +
\om^\l)^2 + 2 - k \bmod 2n. &(5D.13c) \cr}
$$
{}From (5D.12a) and (5D.12b), we see $\si(\omega^\ell)\not=
\la^1.$  Comparing (5D.12a) with (5D.12c), we  obtain either
$\si(\om^\l-e_\l)=\la^2+e_1$ or $\la^2-e_{\l}$. But the former is ruled out by
the norm condition (5D.13b), and the latter leads by (5D.13c) to
$k=2$, contrary to the assumption $k \geq 3$. Thus $\la^2$ must also be
excluded. The weight $\la^3$ is similarly ruled out, because the norm condition
either leads to a contradiction, or else forces $k=2$.

\vskip 0.5truecm \noindent
{\it (iii)}\\
We finish the proof by showing that there is no automorphism satisfying
$\si(\om^1)=J_v(\om^1)$ if both $k\equiv 0 \bmod 4$ and $\l$ odd.

As in step {\it (ii)}, the fusion
$$
\om^1 \times \om^1 = J_v(\om^1) \times J_v(\om^1) = 0 + \psi + (2\om^1)
\eqno(5D.14)
$$
shows that $\psi$ must be fixed, consequently that $\la=\si(\om^\l)$ must be
one
of the twelve weights in (5D.10) (see the argument in step {\it (ii)}).

\smallskip
Assume first $\la=\om^\l$. Using $N_{J_v(\lambda),\mu}^{J_v(\nu)}
= N_{\lambda,\mu}^\nu$ leads to
$$\eqalignno{
& \om^1 \times \om^\l = (\om^\l + e_1) + (\om^\l - e_\l), &(5D.15a) \cr
& J_v(\om^1) \times \om^\l = (\om^\l + (k-1)e_1) + (\om^{\l-1} + (k-2)e_1).
&(5D.15b) \cr}
$$
Trying to match the r.h.s. of (5D.15a) and (5D.15b), the norm forces either $k$
odd or $k=2$. Thus $\la=\om^\l$ is impossible, as is its conjugate
$\la=\om^{\l-1}$.

The next two possibilities, $\la=(k-1,0,\ldots,0,1,0)$ and its conjugate,
correspond to $\la=J_v(\om^\l)$ and $\la=C_1 J_v(\om^\l)$, which require
$k$ odd.

The case $\la=(k-1,0,\ldots,0)=J_v(\om^1)$ is clearly impossible since it is
already the image of $\om^1$. As to $\la=\om^1$, it is forbidden for the same
reason as in step {\it (ii)}, namely because $\si(\om^1 \times \om^\l) =
\om^1 \times J_v(\om^1)$ and that the two fusions do not contain the same
number of fields, see (5D.7c) and (5D.11).

The first and fourth weights of (5D.10c) are ruled out as in step {\it (ii)}.
{}From (5D.12b), we obtain ($\la^1 = (0,\ldots,0,1,k-1)$) $$
J_v(\om^1) \times \la^1 = J_v(\la^1 + e_\l) + J_v(\la^1 - e_{\l-1}) +
J_v(\la^1 - e_\l). \eqno(5D.16)
$$
Since there are only two weights on the r.h.s. of (5D.12a), (5D.16) implies
$\si(\om^\ell)\not=\la^1.$ Similarly, $\si(\om^\ell)\not=(0,\ldots,0,k-1,1).$

There now remain four weights in (5D.10c), namely $\la^2=(0,\ldots,0,k-1)$,
$\la^3 = (1,0,\ldots,0,k-1,0)$, and their $C_1$--conjugates. But the norm
condition implies ${k\l \over 2}\equiv\l \bmod 4$ if $\si(\om^\l)=\la^2$, and
${k\l \over 2}\equiv
\l+2 \bmod 4$ if $\si(\om^\l)=\la^3$, and these congruences are not consistent
with the assumptions $\l$ odd and $k\equiv 0 \bmod 4$ that we made. This
finishes the
proof of step {\it (iii)}, and that of the Theorem for the $D_{\l,k}$
algebras, $k \neq 2$. \cqfd

\vskip 1truecm
\noindent{\bf 6. The Orthogonal Algebras, Level 2}

\vskip 0.7truecm \noindent
As already clear from Table 3, something special happens for the orthogonal
algebras when the level $k$ is equal to 2: a large number of fields have
equal quantum dimensions. This has the immediate consequence that the
technique we used in the previous section is no longer available. More
importantly however it hints at the fact that the current algebras $B_{\l,2}$
and
$D_{\l,2}$ have a much richer spectrum of modular invariants than at the other
levels. Indeed, exceptional automorphism invariants have been recently
discovered in [12] for most orthogonal algebras, level 2. It is the purpose of
this section to show that the list of automorphisms anticipated in [12] form
the complete set, and to give a detailed description of them.

\vskip 0.5truecm \noindent
{\it 6.1. The B--Series, Level 2}

\medskip
The alc\^ove $P_+(B_{\l,2})$ contains $\l+4$ weights: the identity, the $\l$
fundamental weights $\om^i$, and the three combinations $2\om^1$,
$\om^1+\om^\l$ and $2\om^\l$. For what follows, it is convenient to rename $\l$
of these weights as
$$
\nu^i := \om^i, \quad \hbox{for $1 \le i \le \l-1$} \qquad {\rm and} \qquad
\nu^\l := 2\om^\l\,.
\eqno(6.1)
$$
At level 2, the height for $B_{\l,2}$ is equal to $n=2\l+1$.

\smallskip
For any number $x$, we define $[x]_n$ to be the unique number $0\le [x]_n\le {n
\over 2}$ satisfying $x \equiv \pm [x]_n \bmod n$ for some choice of sign.
Then for each integer $m$ satisfying $m^2 \equiv 1 \; \bmod n$, we define the
following permutation of $P_+(B_{\l,2})$
$$
\si_m \;:\;\; \cases{
\si_m(\nu^i) = \nu^{[mi]_n} & for all $1 \le i \le \l$, \cr
\noalign{\smallskip}
\si_m(\la) = \la & if $\la \in \{0,2\om^1,\om^1+\om^\l,\om^\l\}$. \cr}
\eqno(6.2)
$$

We leave the proof that the $\si_m$ actually define automorphism invariants for
section 6.3, where we interpret them as generalized Galois automorphisms.
Note that, since
$\si_m(\nu^1)=\nu^{[m]_n}$,  we obtain that $\si_m = \si_{m'}$ if and only if
$[m]_n=[m']_n$, or equivalently
$m\equiv \pm m' \bmod n$ for some sign. It is easy to show that if $p$ denotes
the number of distinct prime divisors of $n$, then the number of distinct
$\si_m$ is equal to $2^{p-1}$. Also note that $\si_m \circ \si_{m'} =
\si_{mm'}$ so that all automorphisms commute and are of order 2. All but
$\si_1$ are exceptional. We want to show that the $\si_m$ maps are the complete
set of automorphism invariants for $B_{\l,2}$.

\medskip
The quantum dimensions of $B_{\l,2}$ are given in [24]:
$$\eqalignno{
& \D(0) = \D(2\om^1) = 1, &(6.3a) \cr
& \D(\om^1+\om^\l) = \D(\om^\l) = \sqrt{n}, &(6.3b) \cr
& \D(\nu^i) = 2, \quad \hbox{for all $1 \le i \le \l$.} &(6.3c) \cr}
$$
Let $\si$ be any automorphism of $B_{\l,2}$. The eqs (6.3), together with
(3.8a),(3.8c), force $\si(\la) = \la$ for $\la \in
\{0,2\om^1,\om^1+\om^\l,\om^\l\}$. Write $\si(\nu^1)=\nu^m$; the norm
condition (3.8a) then reduces to $n-1 \equiv (n-m)m \bmod 2n$, {\it i.e.}
$m^2 \equiv 1 \bmod n$. Now, $\si$ and
$\si_m$ have the same action on $\om^1$ and $\om^\l$. Thus $\si_m \circ
\si$ leaves them both fixed, and must be the identity by step {\it (i)} of
section 5.2, proving $\si=\si_m$ everywhere. \cqfd

\vskip 0.5truecm \noindent
{\it 6.2. The D--Series, Level 2}

\medskip
The height here is
$n=2+h^\vee=2\ell$. The $\ell+7$ weights in $P_{+}(D_{\ell,2})$ will be
denoted by
$$\eqalignno{
& \ka^1,\,\ka^2,\, \ka^3,\, \ka^4 := 0,\; 2\om^1,\; 2\om^\l,\; 2\om^{\l-1},
&(6.4a) \cr
& \mu^1,\, \mu^2,\, \mu^3,\, \mu^4 := \om^{\l-1},\; \om^\l,\;
\om^1+\om^{\l-1},\; \om^1+\om^\l, &(6.4b) \cr
& \nu^i:=\om^i, \quad \hbox{for $1 \le i \le \l-2$},\quad {\rm and} \quad
\nu^{\ell-1}:=\om^{\ell-1}+\om^\ell. &(6.4c) \cr}
$$

For each $m$ satisfying $m^2 \equiv 1 \bmod 4\ell$, we
define a mapping $\si_m$ on $P_{+}(D_{\ell,2})$ by
$$
\si_m \;:\;\; \cases{
\si_m(\nu^i) = \nu^{[mi]_n} & for all $1 \le i \le \l-1$, \cr
\noalign{\smallskip}
\si_m(\la) = \la & if $\la \in \{\ka^i,\, \mu^i\}_{1 \le i \le 4}$, \cr}
\eqno(6.5a)
$$
with the same definition of $[x]_n$ as in the previous subsection. It will
follow from section 6.3 that all $\si_m$ are generalized Galois automorphisms,
and as such, that they define automorphism invariants.

Our task in this subsection is to prove
the following. For $\l=4$, there are precisely six automorphism
invariants, namely the six conjugations $C_i$ (all $\si_m$ are trivial in this
case). For $\ell>4$, any automorphism invariant of $D_{\ell,2}$
equals $C_1^a\si_m$ for $a=0,1$, and $\si_m$ as in (6.5a). Moreover,
$C_1^a\si_m=C_1^{a'}\si_{m'}$ iff both $a\equiv a'$ mod 2 and $m\equiv \pm
m'$ mod $2\ell$ for some choice of sign.

Thus the
number of automorphism invariants for $\ell>4$ is precisely $2^p$, where
$p$ is the number of distinct prime divisors of $\ell$. When $\ell\not\equiv 2
\bmod 4$, all but two of these, namely $\si_1$ and $C_1$, are exceptional
($\si_s=\si_c=C_1$ and $\si_{vsc}=\si_1$); when $\ell\equiv 2 \bmod 4$,
all but four of them, namely $C_1^a \si_{\ell-1}^b$, are
($\si_s=\si_c=\si_{\l-1}$). Note that for $\l>4$, the composition law is
$$
C_1^a\si_m\circ C_1^{a'}\si_{m'}=C_1^{a+a'}\si_{mm'},
\eqno(6.5b)
$$
so that the automorphisms are all of order 2 and commute.

\medskip
Let us begin with $D_{4,2}$. Computing the norms, we find that $\om^1$,
$\om^3$,
and $\om^4$ are the only weights in the alc\^ove with norm equal to 5 mod
16, and
$\om^2$ is the only one with norm equal to 10 mod 16. Therefore any
automorphism
$\si$ must fix $\om^2$ and permute $\om^1$, $\om^3$, and $\om^4$. Thus for
one of the conjugations $C_i$ of $D_4$,
$C_i\circ\si$ fixes all the $\om^j$, so must equal the identity by Lemma 1.

The quantum dimensions for $D_{\ell,2}$ are computed in [24]:
$$\eqalignno{
& \D(\ka^i) = 1 \qquad \hbox{for $1 \le i \le 4$,} &(6.6a) \cr
& \D(\mu^i) = \sqrt{\l} \qquad \hbox{for $1 \le i \le 4$,} &(6.6b) \cr
& \D(\nu^i) = 2 \qquad \hbox{for $1 \le i \le \l-1$.} &(6.6c) \cr}
$$
For $\ell>4$, $\D(\ka^i) < \D(\nu^j) < \D(\mu^k)$, so that the three sets of
weights must be stable under any $\si$. Computing the norms, we find that
$\si\{\mu^1,\mu^2\}=\{\mu^1,\mu^2\}$, so replacing $\si$ by $C_1\circ\si$ if
necessary, we may assume $\si(\mu^2)=\mu^2$. The mapping
$\si(\nu^1)=\nu^m$ is allowed by the norm condition (3.8a) only if $m$
satisfies $m^2
\equiv 1 \bmod 4\l$; since $\si$ and $\si_m$ coincide on
$\{\om^1,\om^\l\}$, they coincide everywhere. \cqfd

\vskip 0.5truecm \noindent
{\it 6.3. Galois and the Level 2 Exceptionals}

\medskip
It is very tempting to interpret the automorphisms $\si_m$ in (6.2) and
(6.5a) as pure Galois automorphisms, but in fact not all are. For $B_{\ell,2}$,
the Galois group (over $\Bbb Q$) of the extension ${\Bbb Q}(S_{\la,\la'})$ is
contained in $\Z^*_{4n}$. Recall that for a Galois transformation $g_a$ to
define an automorphism invariant, it has to fix the identity, $g_a(0)=0$, and
to
commute with the modular matrix $T$. Leaving the $T$--condition aside for the
moment, let us look at the
other one, and let us suppose that the permutation of the alc\^ove induced by
$g_a$ fixes the identity. If that is so, $g_a$ must leave the quantum
dimensions (6.3) invariant, and in particular $g_a(\sqrt{n})=\sqrt{n}$. It
is a standard result in number theory [19] that $g_a(\sqrt{n})=(n/a)_J
\sqrt{n}$, where $(./.)_J$ is the Jacobi symbol, defined in terms of the
Legendre symbol and the prime decomposition of $a$ by:
$$
\big({n \over a}\big)_J = \prod_p \, \big({n \over p}\big)_L^{k_p}, \qquad
\hbox{for $a=\prod_p p^{k_p}$.}
\eqno(6.7)
$$
We conclude that $g_a$ can define an automorphism invariant only if $a$
satisfies $(n/a)_J=+1$, and one can show that, together with the
$T$--condition, this is also a sufficient condition. It is now an easy matter
to
show that the norm condition alone, which roughly speaking amounts to
$a^2\equiv
1 \bmod n$, is not sufficient to guarantee that $(n/a)_J=+1$. If however
$(n/a)_J=+1$ is satisfied, then $\si_m(\la)=g_a(\la)$ is a Galois automorphism
invariant upon setting $m\equiv a \bmod n$. Similar conclusions apply to
$D_{\ell,2}$: the
$T$--condition and $(q/a)_J \cdot (a/2^t)_J=+1$, where $\l=q\cdot 2^t$ and $q$
odd, are necessary and sufficient conditions for $g_a$ to define a pure Galois
automorphism, which then equals $\si_m$ upon setting $m\equiv a \bmod n$.

However one can show that both (6.2) and (6.5a) have the
generalized Galois form (3.11). Whenever $S_{0,0}^2 \in {\Bbb Q}$
(this is satisfied by $B_{\ell,2}$ and $D_{\ell,2}$ with $S_{0,0}^2=1/4n$ in
both cases), then
$$
g(S_{0,0})=\pm S_{0,0}=\epsilon_g(0)\,S_{g(0),0}
\eqno(6.8)
$$
for any element $g$ of the Galois group,
and therefore $g(0)=J(0)$ for some simple current $J$. For $D_{\l,2}$,
$J=id.$ or $J_v$, because $\D(\om^1) \in {\Bbb Q}$ and (3.13)
imply $Q_J(\om^1) \in\Z$. To commute with $T$, $g_a$ must obey
$a^2\equiv 1 \bmod 2nN$ where
$N=1,2,4$ for $\ell\equiv 0,2,\pm 1 \bmod 4$, respectively. On the other hand,
the Galois group for the orthogonal series is $\Z_{Mn}^*$ where $M=2,4$ when
$\ell$ is even, odd, respectively. Now for any $m$ obeying $m^2\equiv 1$ mod
$n$
(for $B_{\ell,2}$) or mod $2n$ (for $D_{\ell,2}$), it is easy to verify that an
$a\in \Z_{Mn}^*$ can be found such that $a\equiv m \bmod n$ and $g_a$ commutes
with $T$. We want to show $\si_m=\si_{g_a}$, up to a conjugation.

First note that for $B_{\ell,2}$, any automorphism $\si$ must fix $\om^\ell$
(see section 6.1), and for $D_{\ell,2}$, either $\si$ or $C_1\circ\si$
will fix $\om^\ell$ (see section 6.2). Therefore $\si_m(\om^\ell)=C'\circ
\si_{g_a}(\om^\ell)$ for some conjugation $C'$. By step {\it (i)} of sections
5.2 and 5.4, it suffices to show $\si_m(\om^1)=C'\circ\si_{g_a}(\om^1)$. This
can be seen from the following formulas
$$\eqalignno{
{\rm for}\ B_{\ell,2}\; :& \qquad {S_{\om^1,\nu^j} \over S_{0,\nu^j}} =
2\cos({2\pi j \over n}), &(6.9a) \cr
{\rm for}\ D_{\ell,2}\; :& \qquad {S_{\om^1,\nu^j} \over S_{0,\nu^j}} =
2\cos({\pi j \over \ell}). &(6.9b) \cr}
$$
Eq. (6.9a) can be found in [24], while (6.9b) can be derived directly from the
formula
$$
S_{\om^1,\la} = S_{0,\la} \sum_{\mu \in P(\om^1)} \, \exp[-2\pi
i\mu\cdot (\la+\rho)/n],
\eqno(6.10)
$$
where $P(\om^1)$ is the weight diagram of the defining representation of $B_\l$
and $D_\l$. Clearly we have $2 = g_a(\D(\om^1)) = \D(g_a(\om^1))$, so that
$g_a(\om^1)=\nu^j$ for some
$j$. Applying $g_a$ to (6.9a) and (6.9b) yields $g_a(\om^1)=\nu^{[a]_n}$, and
consequently $C' \circ \si_{g_a}(\om^1) = C' \circ J(g_a(\om^1)) =
\nu^{[a]_n}$, as claimed.

So we have proved that, up to conjugation, all automorphisms of $B_{\l,2}$ and
$D_{\l,2}$ are generalized Galois automorphisms of the form (3.11). Let us also
mention that in some cases, all of them can in fact be realized as pure Galois
automorphisms as well as generalized ones. The reason for this is that, to a
single $m$ satisfying $m^2 \equiv 1 \bmod n$ or $2n$, there are in general
several $a \in \Z_{Mn}^*$ such that $a \equiv m \bmod n$ and $g_a$ commutes
with $T$. This happens for instance when $\l$ is odd. For $B_{\l,2}$, $\l$ odd,
$g_a$ commutes with $T$ if $a^2 \equiv 1 \bmod 8n$. But then $a'=a+2n$ also
satisfies $a'^2 \equiv 1 \bmod 8n$, and both $a$ and $a'$ lead to the same $m$
by reduction modulo $n$. They however make a difference because $(n/a')_J =
(-1)^{\l n} (n/a)_J$, so that, for $\l$ odd, either $g_a$ or $g_{a'}$ fixes the
identity. Hence $\si_m=\si_{g_a}$ or $\si_{g_{a'}}$ is a pure Galois
automorphism. The same conclusion holds for $D_{\l,2}$ when
$\l \equiv 3 \bmod 4$.

\vskip 1truecm
\noindent{\bf 7. The Exceptional Algebras}

\vskip 0.7truecm \noindent
We complete in this section the proof of the Theorem for the
five exceptional simple Lie algebras. Fusion rules will be most easily
presented by writing decompositions of tensor products of finite Lie algebra
representations, since fusion
coefficients are identical to the coefficients of truncated
tensor products [8,25], with the truncation related in a simple way to
the level. Explicitly, we can write $$ \la \otimes \la' =
\bigoplus_\mu\bigoplus_{k_t}\; m(k_t)_{\la,\la'}^\mu \;\ \big(\mu\big)_{k_t},
\qquad  \eqno(7.1a) $$
with the fusion coefficient $N_{\la,\la'}^\mu$ at level $k$ determined by
$$N_{\la,\la'}^\mu\ =\ \sum_{k_t=0}^k\ m(k_t)_{\la,\la'}^\mu\ \ .\eqno(7.1b)$$
(7.1a) presents the fusion rules for all levels in an economical way. It
is just
the tensor product $\la\otimes\la',$ with the representations in the
decomposition labelled by the threshold level $k_t$ at and above which the
corresponding affine representation appears in the corresponding fusion rule.
In particular, the tensor product coefficients (or Clebsch-Gordan series
coefficients) are    $$R_{\la,\la'}^\mu\ =\ \sum_{k_t=0}^\infty
m(k_t)_{\la,\la'}^\mu\ \ .\eqno(7.1c)$$  For convenience we will also include
a superscript indicating the ``norm squared'' of the highest weight of the
representations in the tensor product decomposition:
$$ \la \otimes \la' =
\bigoplus_\mu\bigoplus_{k_t}\; m(k_t)_{\la,\la'}^\mu \;\
\big(\mu\big)^{n(\mu)}_{k_t}, \qquad\hbox{with $n(\mu):=(\rho + \mu)^2$.}
\eqno(7.1d) $$

\vskip 0.5truecm \noindent
{\it 7.1. The Algebra $E_6$}

\medskip
The colabels of $E_6$ are equal to $(1,2,3,2,1,2)$, so that a weight of
$P_+(E_{6,k})$ satisfies $\la_0+\la_1+2\la_2+3\la_3+2\la_4+\la_5+2\la_6=k$ and
the dual Coxeter number is equal to 12. The charge conjugation acts as $C(\la)
= (\la_0;\la_5,\la_4,\la_3,\la_2,\la_1,\la_6)$.

There is a simple current $J$ of order 3, given by
$J\la=(\la_5;\la_0,\la_6,\la_3,\la_2,\la_1,\la_4)$. Its charge and
weight are $Q_J(\la)=(-\la_1+\la_2-\la_4+\la_5)/3$ and $h_{J(0)}=2k/3$.
When $k$ is coprime with 3, there is a simple current automorphism invariant
[1]
$$
\si_J(\la)=J^{k(\la_1-\la_2+\la_4-\la_5)}(\la), \qquad \hbox{if $(k,3)=1$.}
\eqno(7.2)
$$
Note that for $k=1,2$, $\si_J=C$.

When $k\equiv 0 \bmod 3$, there are only two automorphism invariants,
$\si_1$ and
$C$, and there are two more when $k \geq 4$ and 3 are coprime, namely
$\si_J$ and $C\si_J$ ($\si_J$ is of order 2, $\si_J^2=\si_1$).

\medskip
Let $\si$ be any automorphism invariant of $E_{6,k}$. From the corollary of
section 4, we have that, for any $k$, $\si(\om^1)=C^a J^b(\om^1)$ for some
$a=0,1$, $b=0,1,2$. Replacing $\si$ with $C^a \circ \si$, we may assume $a=0$.

Consider first $b=1$. The norm condition yields
$$
(\rho + \om^1)^2 \equiv  (\rho + J\om^1)^2 \; \bmod 2n \quad \Rightarrow \quad
{4 \over 3}\,(k-1)\,n \equiv 0 \; \bmod 2n.
\eqno(7.3)
$$
Therefore $\si(\om^1)=J(\om^1)$ requires $k\equiv 1 \bmod 3$. But for precisely
these $k$, the automorphism invariant $\si_J$ has the same action on $\om^1$,
$\si_J(\om^1)=J(\om^1)$. Replacing $\si$ with $\si_J \circ \si$ for these $k$,
we may assume $b=0$. The argument for $b=2$ is identical, so that for all $k$,
we may assume $\si(\om^1)=\om^1$, and prove that the only automorphism with
that property is $\si_1$.

\medskip
Consider first the finite-dimensional Lie algebra tensor product
$$
\om^1 \x \om^1 = \b{\om^2}^{111{1 \over 3}}_2 \;\+\;
\b{\om^5}^{95{1 \over 3}}_1 \;\+\; \b{2\om^1}^{115{1 \over 3}}_2,\eqno(7.4a)
$$
with subscript threshold levels indicating the corresponding fusions.
Since $\om^1$ is fixed by $\si$, the weights on the r.h.s. of (7.4a) must be
permuted. However, from the superscripts, we read that the norms are all
different, so that the permutation must be trivial. In particular $\om^2$
and $\om^5$ must be fixed as well. By considering in a similar way the
following
sequence of tensor products, we can establish that all fundamental weights of
$E_6$
are fixed by $\si$, for all levels $k\ge 1$. The following three tensor
product decompositions are
sufficient:
$$\eqalignno{
& \om^1 \otimes \om^5 = \b{0}^{78}_1 \;\+\; \b{\om^6}^{102}_2 \;\+\;
\b{\om^1+\om^5}^{114}_2 \;, &(7.4b) \cr \noalign{\medskip}
& \om^1 \otimes \om^2 = \b{\om^3}^{126}_3 \;\+\; \b{\om^6}^{102}_2 \;\+\;
\b{\om^1 + \om^2}^{132}_3 \;\+\; \b{\om^1 + \om^5}^{114}_2 \;, &(7.4c) \cr
\noalign{\medskip}
& \om^1 \otimes \om^6 = \b{\om^1}^{95{1 \over 3}}_2 \;\+\;
\b{\om^4}^{111{1 \over 3}}_2 \;\+\; \b{\om^1 + \om^6}^{121{1 \over 3}}_3 \;.
&(7.4d) \cr}
$$
We note that $\om^i$, $i \neq 1$, appears in the fusions (7.4) if and only if
the value of $k$ allows it to be in $P_+(E_{6,k})$. Thus all fundamental
weights in the alc\^ove must be fixed by $\si$, and by Lemma 1,
this shows that $\si=\si_1$ as soon as $\si(\om^1)=\om^1$, and the proof is
complete. \cqfd

\vskip 0.5truecm \noindent
{\it 7.2. The Algebra $E_7$}

\medskip
A weight in the
alc\^ove satisfies $\la_0+2\la_1+3\la_2+4\la_3+3\la_4+2\la_5+\la_6+2\la_7=k$,
and the dual Coxeter number is $h^\vee =18$. The charge conjugation is trivial,
but there is a simple current $J$ of order 2, given by
$J(\la)=(\la_6;\la_5,\ldots,\la_1,\la_0,\la_7)$. It has $Q_J(\la)=(\la_4+
\la_6+\la_7)/2$ and $h_J=3k/4$. When $k\equiv 2 \bmod 4$, it gives rise to
the simple current automorphism invariant [3]
$$
\si_J(\la)=J^{\la_4+\la_6+\la_7}(\la), \qquad \hbox{if $k\equiv 2 \bmod 4$.}
\eqno(7.5)
$$
We note that $\si_J=\si_1$ for $k=2$.

There is only the trivial automorphism $\si=\si_1$ when $k \not\equiv 2 \bmod
4$ or $k=2$, and for $k>2$ and $k\equiv 2 \bmod 4$, there are two, $\si_1$
and $\si_J$.

\medskip
{}From the corollary of section 4, we know that, for any automorphism and any
value of $k$, $\si(\om^6)=J^b(\om^6)$ for some $b=0,1$. Suppose $b=1$. Then
$$
(\rho + \om^6)^2 \equiv (\rho + J\om^6)^2 \; \bmod 2n \quad \Rightarrow \quad
({3 \over 2}k-3)\,n \equiv 0 \; \bmod 2n.
\eqno(7.6)
$$
Thus $\si(\om^6)=J(\om^6)$ requires $k\equiv 2 \bmod 4$. Precisely for these
$k$, the automorphism invariant $\si_J$ has the property that $\si_J(\om^6)=
J(\om^6)$. Replacing $\si$ with $\si_J \circ \si$, we may assume $b=0$.
Thus for all $k$, we may assume $\si(\om^6)=\om^6$, and prove that the only
such
automorphism is $\si_1$.

\smallskip
For $k \geq 1$, all fundamental weights $\om^i$ belonging to $P_+(E_{7,k})$
appear in the fusions (7.7) and the usual argument shows that they must be
fixed by $\si$. The fusion threshold levels of the following tensor
products [33] were obtained using the affine Weyl group:
$$\eqalignno{
& \om^6 \otimes \om^6 = \b{0}^{199.5}_1 \;\+\; \b{\om^1}^{235.5}_2 \;\+\;
\b{\om^5}^{255.5}_2 \;\+\; \b{2\om^6}^{259.5}_2\;, &(7.7a) \cr
\noalign{\medskip}
& \om^1 \otimes \om^6 = \b{\om^6}^{228}_2 \;\+\; \b{\om^7}^{252}_2 \;\+\;
\b{\om^1 + \om^6}^{266}_3\;, &(7.7b) \cr
\noalign{\medskip}
& \om^1 \otimes \om^1 = \b{0}^{199.5}_2 \;\+\; \b{\om^1}^{235.5}_3 \;\+\;
\b{\om^2}^{271.5}_3 \;\+\; \b{\om^5}^{255.5}_2 \;\+\; \b{2\om^1}^{275.5}_4\;,
&(7.7c) \cr \noalign{\medskip}
& \om^1 \otimes \om^7 = \b{\om^4}^{282}_3 \;\+\; \b{\om^6}^{228}_2 \;\+\;
\b{\om^7}^{252}_3 \;\+\; \b{\om^1 + \om^6}^{266}_3 \;\+\;
\b{\om^1 + \om^7}^{292}_4\;,\qquad &(7.7d) \cr \noalign{\medskip}
& \om^7 \otimes \om^7 = \b{0}^{199.5}_2 \;\+\; \b{\om^1}^{235.5}_3 \;\+\;
\b{\om^2}^{271.5}_3 \;\+\; \b{\om^3}^{307.5}_4 \;\+\; \b{\om^5}^{255.5}_3
\;\+\; \b{2\om^1}^{275.5}_4 \cr
\noalign{\medskip}
& \qquad \qquad \+\; \b{2\om^6}^{259.5}_2 \;\+\;
\b{2\om^7}^{311.5}_4 \;\+\; \b{\om^1 + \om^5}^{295.5}_4 \;\+\;
\b{\om^6 + \om^7}^{283.5}_3\;. &(7.7e) \cr}
$$
Going through these five products and assuming $\si(\om^6)=\om^6$, we have
successively that
$\om^1,\,\om^5,\,\om^7,\,\om^2,\,\om^4$ and $\om^3$ must be fixed by
$\si$. By Lemma 1, the whole of the alc\^ove must be fixed. \cqfd

\vskip 0.5truecm \noindent
{\it 7.3. The Algebra $E_8$}

\medskip
Here a weight satisfies
$\la_0+2\la_1+3\la_2+4\la_3+5\la_4+6\la_5+4\la_6+2\la_7+3\la_8=k$, and the dual
Coxeter number is 30. The charge conjugation is trivial and there is no simple
current (except for an anomalous one at $k=2$ which does not give rise to an
automorphism invariant).

\smallskip
We will show that for all levels $k \neq 4$, there is only the trivial
automorphism invariant $\si_1$, and that for $k=4$, there is a second,
exceptional one we call $\si_{e8}$, and which was first given in [10]. It
permutes the fundamental weights $\om^1$ and $\om^6$ and fixes all other
weights:
$$
\si_{e8} \;:\;\; \cases{ \om^1 \; \longleftrightarrow \; \om^6, & \cr
\hbox{fixes all other weights.} & \cr}
\eqno(7.8)
$$
This is {\it not} a Galois automorphism: the Galois group for $E_{8,k}$ is
$\Z_{k+30}^*$; $g_a$ commutes with $T$ iff $a^2\equiv 1$ mod $n$; so the
only possible $g_a$ at $k=4$ are $g_1$ and $g_{-1}$, both of which give
$\si_1$. Remarkably, in the set of all automorphism invariants for all simple
$X_\l$ and all levels $k$, $\si_{e8}$ is the {\it only one} that cannot be
explained in terms of simple currents, conjugations, Galois transformations or
these combined.

\medskip
Let $\si$ be any automorphism invariant of $E_{8,k}$. From the corollary of
section 4, we have that, for any $k\ne 4$, $\si(\om^1)=\om^1$. For $k=4$, the
only other weight in the alc\^ove which has the same quantum dimension as
$\om^1$ is $\om^6$, so that the additional possibility is $\si(\om^1)=\om^6$.
But in this case we can replace $\si$ with $\si_{e8} \circ \si$ so that the new
$\si$ fixes $\om^1$. Thus for all $k$, we may assume $\si(\om^1)=\om^1$. The
proof will be complete if we show that any such automorphism is necessarily
trivial.

\smallskip
We show that if $\om^1$ is fixed, then so are all $\om^i$, for $1\le i\le 8$,
which are in the alc\^ove. By
the usual norm argument, the fusions encoded in the following sequence of
tensor products establishes the result {\it except} for $\om^5$:
$$\eqalignno{
& \om^1 \otimes \om^1 = \b{0}^{620}_2 \;\+\; \b{\om^1}^{680}_3 \;\+\;
\b{\om^2}^{740}_3 \;\+\; \b{\om^7}^{716}_2 \;\+\; \b{2\om^1}^{744}_4\;, &(7.9a)
\cr \noalign{\medskip}
& \om^1 \otimes \om^7 = \b{\om^1}^{680}_2 \;\+\; \b{\om^2}^{740}_3 \;\+\;
\b{\om^7}^{716}_3 \;\+\; \b{\om^8}^{764}_3 \;\+\; \b{\om^1+\om^7}^{780}_4\;,
&(7.9b) \cr \noalign{\medskip}
& \om^1 \otimes \om^8 = \b{\om^2}^{740}_3 \;\+\; \b{\om^3}^{800}_4 \;\+\;
\b{\om^6}^{816}_4 \;\+\; \b{\om^7}^{716}_3 \;\+\; \b{\om^8}^{764}_4 \cr
\noalign{\medskip}
& \qquad \qquad \+\; \b{\om^1+\om^7}^{780}_4 \;\+\;
\b{\om^1+\om^8}^{830}_5\;, &(7.9c) \cr \noalign{\medskip}
& \om^1 \otimes \om^3 = \b{\om^2}^{740}_4 \;\+\; \b{\om^3}^{800}_5 \;\+\;
\b{\om^4}^{860}_5 \;\+\; \b{\om^6}^{816}_4 \;\+\; \b{\om^8}^{764}_4 \;\+\;
\b{\om^1+\om^2}^{806}_5 \cr \noalign{\medskip}
& \qquad \qquad \+\;\b{\om^1+\om^3}^{868}_6 \;\+\; \b{\om^1+\om^7}^{780}_4
\;\+\;
\b{\om^1+\om^8}^{830}_5 \;\+\; \b{\om^2+\om^7}^{844}_5 \;. &(7.9d) \cr}
$$
These fusions were calculated from the corresponding tensor products listed in
[26] using the affine Weyl group.

The remaining fundamental representation, $\om^5$, is contained in
the following tensor product:
$$\eqalignno{
& \om^6 \otimes \om^7 = \b{\om^2}^{740}_4 \;\+\; \b{\om^3}^{800}_5 \;\+\;
\b{\om^4}^{860}_5 \;\+\; \b{\om^5}^{920}_6 \;\+\; 2\b{\om^6}^{816}_{4,5} \;\+\;
\b{\om^7}^{716}_4 \cr \noalign{\medskip}
& \quad \quad \+\; \b{\om^8}^{764}_4 \;\+\; \b{\om^1+\om^2}^{806}_5 \;\+\;
\b{\om^1+\om^3}^{868}_6 \;\+\; \b{\om^1+\om^6}^{884}_6 \;\+\;
2\b{\om^1+\om^7}^{780}_{4,5} \cr \noalign{\medskip}
& \quad \quad\+\; 2\b{\om^1+\om^8}^{830}_{5,5} \;\+\;
2\b{\om^2+\om^7}^{844}_{5,5}
\;\+\; \b{\om^2+\om^8}^{896}_6 \;\+\; \b{\om^3+\om^7}^{908}_6 \;\+\;
\b{\om^6+\om^7}^{926}_6 \cr \noalign{\medskip}
& \quad \quad \+\; 2\b{\om^7+\om^8}^{870}_{5,5} \;\+\; \b{2\om^7}^{820}_5
\;\+\;
\b{\om^1+2\om^7}^{888}_6 \;\+\; \b{2\om^1+\om^7}^{848}_6\;. &(7.9e) \cr}
$$
This time, the norm argument is not sufficient to show from (7.9e) that $\si$
must also fix $\om^5$. However, one can show that the only representations that
can possibly be exchanged with $\om^5$ are $\b{2\om^7}^{820}_5$ and
$\b{2\om^1+\om^7}^{848}_6$, and that can only happen for levels $k= 20$
and $k=6$, respectively. But it is then easily checked for these levels that
their quantum dimensions differ, so that at the end
$\om^5$ too must be fixed. Thus, for all $k \geq 2$, all fundamentals $\om^i$
in the alc\^ove must be fixed if $\om^1$ is fixed, and Lemma 1 implies once
more
that $\si=\si_1$, completing the proof for $E_8$. \cqfd

\vskip 0.5truecm \noindent
{\it 7.4. The Algebra $F_4$}

\medskip
A weight in $P_+(F_{4,k})$ satisfies $\la_0+2\la_1+3\la_2+2\la_3+\la_4=k$, and
the dual Coxeter number is $h^\vee=9$. The charge conjugation $C$ is
trivial, and there is no simple current.

We will show that for $k \neq 3$, $\si_1$ is the only automorphism invariant,
and that at $k=3$, there is one more, namely the exceptional $\si_{f4}$, first
found in [35]. It is given by
$$
\si_{f4} \;:\;\; \cases{
{\rm permutes} \; \om^2 \longleftrightarrow \om^4 \quad {\rm and} \quad
\om^1 \longleftrightarrow 3\om^4, &\cr
\hbox{fixes all other weights.} & \cr}
\eqno(7.10)
$$
In fact this permutation equals the one induced by the Galois transformation
$g_5$, given by (3.11) (a pure Galois automorphism). For $k=3$, the relevant
Galois group is isomorphic to $\Z_{24}^*$, and one finds, in the notation of
section 3, that $g_5(\la)=\si_{f4}(\la)$.

\smallskip
{}From the corollary of section 4, we have that for $k \neq 3$, an automorphism
must satisfy $\si(\om^4)=\om^4$. From Table 3, the only other possibility at
$k=3$ is $\si(\om^4)=\om^2$, but in this case, we can replace $\si$ by
$\si_{f4} \circ \si$ and assume that $\om^4$ is fixed. The conclusion follows
if we show that $\si(\om^4)=\om^4$ implies $\si=\si_1$.

This is easily done with the following two tensor products
$$\eqalignno{
& \om^4 \otimes \om^4 = \b{0}^{39}_1 \;\+\; \b{\om^1}^{57}_2 \;\+\;
\b{\om^3}^{63}_2 \;\+\; \b{\om^4}^{51}_1 \;\+\; \b{2\om^4}^{65}_2 \;, &(7.11a)
\cr \noalign{\medskip}
& \om^3 \otimes \om^4 = \b{\om^1}^{57}_2 \;\+\; \b{\om^2}^{75}_3 \;\+\;
\b{\om^3}^{63}_2 \;\+\; \b{\om^4}^{51}_2  \cr \noalign{\medskip}
& \qquad \qquad \qquad \+\; \b{\om^1+\om^4}^{71}_3 \;\+\;
\b{\om^3+\om^4}^{78}_3
\;\+\; \b{2\om^4}^{65}_2 \;. &(7.11b) \cr}
$$
The norm condition implies that all representations in these two products must
be fixed by $\si$ if $\om^4$ is fixed, so in particular those fundamentals
lying in the alc\^ove are fixed, implying $\si=\si_1$ by Lemma 1. \cqfd

\vskip 0.5truecm \noindent
{\it 7.5. The Algebra $G_2$}

\medskip
A weight in the alc\^ove satisfies $\la_0+2\la_1+\la_2=k$, and $h^\vee=4$.
There
is no charge conjugation nor simple current.

The only non--trivial automorphism invariant $\si_{g2}$ is found at level $k=4$
[35]. It is the following permutation
$$
\si_{g2} \;:\;\; \cases{
{\rm permutes} \; \om^1 \longleftrightarrow 4\om^2 \quad {\rm and} \quad
2\om^1 \longleftrightarrow \om^2, &\cr
\hbox{fixes all other weights.} & \cr}
\eqno(7.12)
$$
The Galois group $\Z_{24}^*$ is the same as for $F_{4,3}$, and $g_5(\la) =
\si_{g2}(\la)$ also holds here.

\smallskip
{}From the corollary of section 4, the second fundamental weight $\om^2$ must
be
left invariant by any $\si$, for $k \neq 4$. At $k=4$, Table 3 shows that the
only other possibility is $\si(\om^2)=2\om^1$; in this case, composing $\si$
with $\si_{g2}$ allows to assume that, here too, $\om^2$ must be fixed.

It remains to show that an automorphism fixing $\om^2$ is trivial. This
immediately follows from the tensor product
$$
\om^2 \otimes \om^2 = \b{0}^{4{2 \over 3}}_1 \;\+\; \b{\om^1}^{12{2 \over 3}}_2
\;\+\; \b{\om^2}^{8{2 \over 3}}_1 \;\+\; \b{2\om^2}^{14}_2\,,
\eqno(7.13)
$$
which shows that $\om^1$ is fixed, and from Lemma 1. \cqfd

\vskip 1truecm
\noindent{\bf 8. Conclusion}

\vskip 0.7truecm \noindent
In this paper, we have established the complete list of automorphism modular
invariants for unextended current models based on finite--dimensional simple
Lie algebras. More precisely,
we have classified the modular invariant forms, sesquilinear in the affine
characters, obtained by twisting the diagonal invariant by an automorphism
 of the fusion ring. Some of these invariants correspond to
torus partition functions of WZW models
[13]. In particular, the diagonal invariants describe WZW models based on
simply-connected simple Lie groups. The WZW partition functions for
nonsimply-connected groups can be obtained by ``orbifolding'' the diagonal
invariant [18]. Many of our invariants can be obtained by similarly
``orbifolding'' the conjugation invariant. But many others await
interpretation. For example, those Galois automorphism
invariants which cannot be written as simple current invariants
seem problematic at present.  Another interesting case is provided by the
exceptional invariant of $E_{8,4}$ which, in the list of invariants for all
simple algebras and all levels, is the only one that cannot be described in
terms of simple currents, conjugations and/or Galois transformations.

Although the list of automorphism invariants constitutes major progress
towards the
general problem of classifying modular invariants for conformal current
models, more technical problems
need to be overcome before the full list of modular invariants can be
contemplated. Among the most striking ones, let us mention the fixed point
resolution problem, and more importantly, the classification of the chiral
extensions of Kac--Moody algebras. A humbler task should be to extend our
analysis to the remaining semi-simple Lie algebras --- something of direct
value for the Goddard--Kent--Olive models.

\vskip 1truecm
\noindent{\it Acknowledgments.} T.\ G.\ thanks the IHES and the Concordia
math department for their generosity.
P. R. is grateful to the Departamento de F\'\i sica de Part\'\i culas
(Santiago de Compostela) for the kind hospitality, and acknowledges the
financial support of the EEC Human Capital and Mobility Program, EEC network
on {\it Physical and Mathematical Aspects of the Theory of Fundamental
Interactions} (contract ERBCHRXCT920035). M. W. acknowledges a very helpful
visit at the IHES.
\vfill \eject
\noindent{\bf References.}

\vskip 0.7truecm

\item{1.} Altschuler, D., Lacki, J., Zaugg, Ph.: The affine Weyl group
and modular invariant partition functions. Phys.\ Lett.\ {\bf B205},
281-284 (1988)

\item{2.} Belavin, A.A., Polyakov, A.M., Zamolodchikov, A.B.: Infinite
conformal symmetry in two-dimensional quantum field theory. Nucl. Phys. {\bf
B241}, 333-380 (1984)

\item{3.} Bernard, D.: String characters from Kac-Moody automorphisms.
Nucl. Phys. {\bf B288}, 628-648 (1987)

\item{4.} Bourbaki, N.: Groupes et alg\`ebres de Lie, Chapitres IV-VI.
Paris: Hermann 1968

\item{5.} Cappelli, A., Itzykson, C., Zuber, J.-B.: The A-D-E
classification of $A_1^{(1)}$ and minimal conformal field theories. Commun.
Math. Phys. {\bf 113}, 1-26 (1987)

\item{6.} Cardy, J.: Effect of boundary conditions on the operator content
of conformally invariant theories. Nucl. Phys. {\bf B275}, 200-218 (1986)

\item{7.} Coste, A., Gannon, T.: Remarks on Galois symmetry in RCFT.
Phys.\ Lett.\ {\bf B323}, 316-321 (1994)

\item{8.} Cummins, C., Mathieu, P., Walton, M.\ A.: Generating functions
for WZNW fusion rules. Phys.\ Lett.\ {\bf B254}, 386-390 (1991)

\item{9.} Fuchs, J.: Simple WZW currents. Commun.\ Math.\ Phys.\ {\bf 136},
 345-356 (1991)

\item{10.} Fuchs, J.: WZW Quantum dimensions. Int.\ J.\ Mod.\ Phys.\
{\bf B6}, 1951-1965 (1992)

\item{11.} Fuchs, J., Gato-Rivera, B., Schellekens, A.\ N., Schweigert, C.:
Modular invariants and fusion rule automorphisms from Galois theory.
Phys.\ Lett.\ {\bf B334}, 113-120 (1994)

\item{12.} Fuchs, J., Schellekens, A.\ N., Schweigert, C.: Galois
modular invariants of WZW models. (hepth/9410010)

\item{13.} Felder, G., Gawedzki, K., Kupiainen, A.: Spectra of
Wess-Zumino-Witten models with arbitrary simple groups. Commun.\ Math.\
Phys.\ {\bf 117}, 127-158 (1988)

\item{14.} Furlan, P., Ganchev, A., Petkova, V.: Quantum groups and fusion
rules multiplicities. Nucl.\ Phys.\ {\bf B343}, 205-227 (1990)

\item{15.} Gannon, T.: WZW commutants, lattices, and level-one partition
functions. Nucl.\ Phys.\ {\bf B396}, 708-736 (1993)

\item{16.} Gannon, T.: The classification of affine $su(3)$ modular
invariant partition functions. Commun. Math. Phys. {\bf 161}, 233-264 (1994)

\item{17.} Gannon, T.: Symmetries of the Kac-Peterson modular matrices
of affine algebras. IHES preprint P/94/53 (q-alg/9502004)

\item{18.} Gepner, D., Witten, E.: Strings on group manifolds. Nucl. Phys.
{\bf B278}, 493 (1986)

\item{19.} Hua Loo Keng: Introduction to Number Theory. Berlin:
Springer-Verlag 1982

\item{20.} Intriligator, K.: Bonus symmetry in rational conformal field
theory. Nucl. Phys. {\bf B332}, 541-565 (1990)

\item{21.} Itzykson, C.: Level-one Kac-Moody characters and modular
invariance. Nucl.\ Phys.\ (Proc.\ Suppl.) {\bf B5}, 150-165 (1988);

\item{} Degiovanni, P.: Z/NZ conformal field theories. Commun.\ Math.\
Phys.\ {\bf 127}, 71-99 (1990)

\item{22.} Kac, V.\ G.: Infinite dimensional Lie algebras, 3rd edition
Cambridge: Cambridge University Press 1990

\item{23.} Kac, V.\ G., Peterson, D.: Infinite-dimensional Lie algebras,
theta functions and modular forms. Adv.\ Math.\ {\bf 53}, 125-264 (1984)

\item{24.} Kac, V.\ G., Wakimoto, M.: Modular and conformal invariance
constraints in representation theory of affine algebras. Adv.\ Math.\
{\bf 70}, 156-236 (1988)

\item{25.} Kirillov, A.\ N., Mathieu, P., S\'en\'echal, D., Walton,
M.\ A.: Can fusion coefficients be calculated from the depth rule?
Nucl.\ Phys.\ {\bf B391}, 651-674 (1993)

\item{26.} McKay, W.\ G., Moody, R.\ V., Patera, J.: Decomposition of
tensor products of $E_8$ representations. Algebras, groups and geometries
{\bf 3}, 286-328 (1986)

\item{27.} Moore, G., Seiberg, N.: Naturality in conformal field theory.
Nucl. Phys. {\bf B313}, 16-40 (1989)

\item{28.} Naculich, S.\ G., Riggs, H.\ A., Schnitzer, H.\ J.:
Group-level duality in WZW models and Chern-Simons theory. Phys.\ Lett.\
{\bf B246}, 417-422 (1990);

\item{} Mlawer, E.\ J., Naculich, S.\ G., Riggs, H.\ A., Schnitzer, H.\ J.:
Group-level duality of WZW fusion coefficients and Chern-Simons link
observables. Nucl.\ Phys.\ {\bf B352}, 863-896 (1991)

\item{29.} Ruelle, P., Thiran, E., Weyers, J.: Implications of an
arithmetical symmetry of the commutant for modular invariants. Nucl.\
Phys.\ {\bf B402}, 693-708 (1993);

\item{} Ruelle, P.: Automorphisms of the affine $SU(3)$ fusion rules.
Commun. Math. Phys. {\bf 160}, 475-492 (1994)

\item{30.} Schellekens, A.\ N.: Fusion rule automorphisms from
integer spin simple currents. Phys.\ Lett.\ {\bf B244}, 255-260 (1990)

\item{31.} Schellekens, A.\ N., Yankielowicz, S.: Extended chiral
algebras and modular invariant partition functions. Nucl.\ Phys.\ {\bf
327}, 673-703 (1989)

\item{32.} Schellekens, A.\ N.\ Yankielowicz, S.: Modular invariants
from simple currents: an explicit proof. Phys.\ Lett.\ {\bf B227},
387-391 (1989)

\item{33.} Slansky, R.: Group theory for unified model building.
Phys.\ Rep.\ {\bf 79}, 1 (1981)

\item{34.} Verlinde, E.: Fusion rules and modular transformations in
2D conformal field theory. Nucl.\ Phys.\ {\bf 300 [FS22]}, 360-376 (1988)

\item{35.} Verstegen, D.: New exceptional modular invariant partition
functions for simple Kac-Moody algebras. Nucl.\ Phys.\ {\bf B346},
349-386 (1990)

\item{36.} Walton, M.A.: Algorithm for WZW fusion rules: a proof. Phys.
Lett. {\bf 241B}, 365-368 (1990); Fusion rules in Wess-Zumino-Witten models.
Nucl. Phys. {\bf B340}, 777-790 (1990)

\end